\documentclass[aps,prd,twocolumn,showpacs,groupedaddress,nofootinbib]{revtex4}

\usepackage{graphicx}  % needed for figures
\usepackage{dcolumn}   % needed for some tables
\usepackage{bm}        % for math
\usepackage[english]{babel}
\usepackage{amsfonts,amsmath,amssymb,mathrsfs}
\usepackage{times}

\usepackage{lineno}

\newcommand{\araa}{ARAA}
\newcommand{\aap}{A{\&}A}
\newcommand{\myapjl}{Astrophys. J. Lett.}
\newcommand{\myapj}{Astrophys. J.}
\newcommand{\pasp}{Publ. Astron. Soc. Pac.}
\newcommand{\cqg}{Class. Quant. Grav.}
\newcommand{\mnras}{Mon. Not. Roy. Astron. Soc.}
\newcommand{\myprd}{Phys. Rev. D}
\newcommand{\myprl}{Phys. Rev. Lett.}
\newcommand{\eq}{\begin{equation}}
\newcommand{\eeq}{\end{equation}}
\newcommand\T{\rule{0pt}{3ex}}
\newcommand\B{\rule[-2ex]{0pt}{0pt}}

\def\one{\hbox{$_{1}$}}
\def\two{\hbox{$_{2}$}}

\begin{document}
\title{Gravitational waves from Extreme Mass Ratio Inspirals in
  non-pure Kerr spacetimes}
\author{Enrico Barausse$^{1}$}
\author{Luciano Rezzolla$^{2,3}$}
\author{David Petroff$^{4}$}
\author{Marcus Ansorg$^{2}$}

\affiliation{$^{1}$SISSA, International School for
             Advanced Studies and INFN, Trieste, Italy}

\affiliation{$^{2}$Max-Planck-Institut f\"ur Gravitationsphysik,
             Albert-Einstein-Institut, Potsdam-Golm, Germany}

\affiliation{$^{3}$Department of Physics, Louisiana State University, Baton
             Rouge, USA}

 \affiliation{$^{4}$Theoretisch-Physikalisches Institut, University of Jena,
           Jena, Germany}

\date{\today}
\begin{abstract}
  To investigate the imprint on the gravitational-wave emission from
  extreme mass-ratio inspirals in non-pure Kerr spacetimes, we have studied
  the ``kludge'' waveforms generated in highly-accurate, 
  numerically-generated spacetimes containing a black hole and a
  self-gravitating, homogeneous torus with comparable mass and spin. In order to
  maximize their impact on the produced waveforms, we have considered
  tori that are compact, massive and close to the central black hole,
  investigating under what conditions the LISA experiment could detect
  their presence. Our results show that for a large portion of the
  space of parameters the waveforms produced by EMRIs in these black
  hole-torus systems are indistinguishable from pure-Kerr waveforms.
  Hence, a ``confusion problem'' will be present for observations
  carried out over a timescale below or comparable to the dephasing time.
\end{abstract}

\maketitle

%-----------------------------------------------
\section{\label{sec:intro}Introduction}
%-----------------------------------------------

Extreme Mass Ratio Inspirals (EMRIs) are thought to be one of the most
interesting sources of gravitational waves for the space-based
gravitational-wave detector LISA~\cite{LISA}: the typical example is a
black hole with mass $\sim 1$--$10\, M_{\odot}$ orbiting around the
supermassive black hole (SMBH) at the center of a galaxy. It is
expected that LISA will be able to detect anywhere from tens up to a thousand
of these sources during its lifetime, which will probably be between 3
and 5 years. Although the masses of SMBHs range from $10^6$ to
$10^{10} M_{\odot}$~\cite{SMBH}, the mass of the SMBH involved in an
EMRI must be around $10^6 M_\odot$ in order for the gravitational wave
signal to be within LISA's sensitivity band: see, for instance,
Ref.~\cite{gair_event_rates} for the expected event rates for
different masses of the stellar black hole and of the SMBH.

As in the case of Earth-based detectors, for which the signal is generally
expected to be comparable with the noise, the detection of
gravitational waves emitted by EMRIs and the subsequent
characterization of the source is expected to take place at small
values of the signal-to-noise ratio (SNR), thus requiring some sort of
matched filtering. This method is based on cross-correlating the noisy
gravitational wave signal with a bank of templates, which should accurately
model the true signal, and poses serious challenges both in
building the templates and in accessing them (see
Ref.~\cite{gair_event_rates} for a detailed discussion).

The SMBHs involved in EMRIs are commonly thought to be describable by
the pure Kerr solution of General Relativity: this is the common
assumption made in most work on EMRIs. Nevertheless, a number of
other ``exotic'' candidates have been proposed as alternatives to the
central massive object. These are, for instance,
gravastars~\cite{gravastars}, boson stars~\cite{boson_stars}, fermion
balls~\cite{fermion_balls}, oscillating axion
bubbles~\cite{axionbubble}, etc. Clearly, while it is not yet possible
to exclude completely these possibilities, the presence of these
objects at the centres of galaxies would require a serious modification
to the scenarios through which galaxies are expected to form. At the
same time, the possibility that LISA observations could be used to
determine the presence (or absence) of these objects, provides
additional scientific value to this challenging experiment.

Hereafter, we will adopt a more conservative view and assume that the
central object is indeed an SMBH. Recent observations of the
near-infrared fluxes of SgrA* support this view by setting upper
limits on the mass accretion rate of the Galactic center and showing
that the central massive object must have, under reasonable
assumptions, an event horizon~\cite{broderick_and_narayan}. Yet, even
with this assumption, the modelling of EMRIs can in principle suffer
from the uncertainty of whether the spacetime in the vicinity of the
SMBH can be accurately described in terms of a (pure) Kerr solution.
The origin of this uncertainty is that SMBHs are not expected to be in
vacuum and indeed a considerable amount of matter is expected to be
present around the central massive object. In the case of active
galactic nuclei (AGNs), for instance, the intense high-energy emission
is thought to be the result of a pc-scale accretion disk (and perhaps a thick
torus) extending down almost to the innermost stable circular orbit
(ISCO), feeding the central black hole. In addition, a dusty
obscuration torus is also believed to be present on much larger scales
(i.e., $\sim10$--$100$ pc)~\cite{agn}. Too little is presently known
about the properties of these disks and although their mass is
commonly thought to be much smaller than the mass of the SMBH, there
are observations hinting at disks as massive as the central
object~\cite{non_kelperian}.

Another example is given by SgrA* itself, where counter-rotating
stellar disks on scales less than 1 pc have been
observed~\cite{genzel}. This is hardly surprising since the Galactic
center is expected to be a high density environment, as the
distribution of stars shows a cusp there: the mass density in stars is
believed to be $\rho\sim\rho_0(r/r_0)^{-\alpha}$, with
$\rho_0\sim1.2\times10^6M_\odot/\mbox{pc}^3$, $r_0\sim0.04\mbox{ pc}$
and $\alpha\sim1.4$--$2$~\cite{genzel}.

Furthermore, even if an SMBH exists, it is still possible it could be
surrounded by other, non-visible components, such as clusters of
compact objects or high concentrations of exotic
particles. Cosmological N-body simulations predict, in fact, that the
cold dark matter (CDM) density in galactic halos should show a
``cusp'' near the galactic center with a profile of the
type~\cite{NFW} $\rho_{_{\rm CDM}}\sim\rho_{0}(r/r_0)^{-\alpha}$,
where $\alpha\sim1$: although the mass in CDM particles is generally
thought to be smaller than that in the stellar cusp
($\rho_{0}\sim100M_\odot/\mbox{pc}^2$ and $r_{0}\sim3\mbox{
  pc}$~\cite{DM_cusp}), the normalization of this power law is still
very uncertain.\footnote{The possibility that a steeper profile
  (i.e., larger $\alpha$) could form under the influence of
  the SMBH was also proposed in Ref.~\cite{silk}, although the process
  does not seem to happen in a more realistic astrophysical
  scenario~\cite{ullio}.}  In addition, although this CDM distribution
is commonly thought to be spherically symmetric, the confrontation
with observations still leaves a number of uncertainties, with the
presence of the CDM cusp itself being in contrast with observations of
galactic rotation curves, which instead hint at a CDM core-profile in
galactic centers~\cite{cusp_problem}. The possibility that
CDM could be distributed along caustic rings in galactic halos has
also been suggested~\cite{sikivie}.

Clearly, gravitational-wave observations through the LISA detector
could shed some light on these issues, enabling the distinction between
competing models for the central massive object and for the
distribution of matter around it. Indeed, observations of EMRIs by
LISA could allow us to build a map of the spacetime around 
galactic centers and determine with great precision the properties of
the spacetime in regions which are not easily accessible through
electromagnetic observations.

A number of different approaches to this ``spacetime-mapping'' problem
were considered in the the literature: EMRIs have been studied in
spacetimes which are either approximate or exact solutions of the
Einstein equations.  Among the former, a multipolar expansion suitable
to describe general stationary, axisymmetric, asympotically flat
spacetimes outside a central distribution of matter has been
considered~\cite{ryan_multipoles,ryan_3,ryan_4}. However,
this multipolar expansion is in practice a series in $1/r$ ($r$ being the distance
to the central object) around a Minkowski spacetime: an accurate
representation of the strong field regime would require the inclusion
of many terms. Another possibility is the ``quasi-Kerr''
(i.e., Kerr plus a small quadrupole) spacetime studied by
Glampedakis and Babak~\cite{quasi_kerr}. This can approximately
describe the spacetime outside a slowly rotating boson star and is
\textit{not} an expansion around Minkowski, thus being more
promising in the strong field limit.  Among exact solutions of the
Einstein equations, only spherical boson stars~\cite{gair_boson_star}
and ``bumpy black holes''~\cite{bumpyBHs} (i.e., objects
that, although involving naked singularities, are \textit{almost}
Schwarzschild black holes, but have some multipoles with the wrong
values) have been considered.

At any rate, none of these spacetimes, neither exact nor approximate,
can describe satisfactorily the ``astrophysical bumpiness'' which is
certainly present around SMBHs.  With this in mind, we have studied
EMRIs in stationary, axisymmetric spacetimes which are highly accurate
numerical solutions of the Einstein equations and contain a rotating
black hole and a torus~\cite{BH_plus_ring}.

We used these numerical spacetimes to perform a study similar to that carried out by Babak
and Glampedakis for ``quasi-Kerr'' spacetimes~\cite{quasi_kerr}: we studied EMRIs in the equatorial
plane and computed semi-relativistic (``kludge'') waveforms, 
comparing them to kludge, pure-Kerr waveforms.  Babak and Glampedakis,
in particular, find there could be a ``confusion'' problem, because
although gravitational waves emitted in a quasi-Kerr spacetime by a
stellar mass black hole moving on an equatorial orbit are wildly
different from those emitted by the same stellar mass black hole
moving along the same orbit in a pure Kerr spacetime (having the same
mass and spin as the quasi-Kerr spacetime), waveforms produced by
equatorial orbits having slightly different latus rectum and
eccentricity but the same $r$- and $\phi$-frequencies turn out
to be indistinguishable with LISA's sensitivity. We therefore repeated
and extended their analysis. In particular, we introduce, like them, a
suitable cut-off in time in order not to have any relevant
radiation-reaction effects on the geodetic motion. While this could be
avoided in Babak's and Glampedakis' quasi-Kerr spacetimes (probably
eliminating the confusion problem: see the analysis in
Ref.~\cite{quasi_kerr_cutler}), this is actually a necessity in our
case, since the effect of a torus on the loss of energy and angular
momentum due to gravitational-wave emission is completely unknown at
present.

We did not try, for the moment, to produce tori describing
the accretion disk of AGNs (although we plan to do this in a future
paper), but rather adopted a more phenomenological approach. Indeed,
since little is known about the strong field region near the central
massive black hole, we tried to build some ``extreme'' configurations,
i.e. configurations containing rather massive and compact tori
(close to the event horizon of the central black hole). The
purpose is to understand if LISA can detect the presence of such tori,
which are so close to the horizon that they could not probably be
detected otherwise (for instance, by means of stellar orbits),
especially if made of some ``dark'' mass.  We stress that the word
``extreme'' does not mean that these configurations are extremely far
from Kerr, but just that these tori are \textit{not} the ones
astrophysicists expect in AGNs.

One possible objection is that it might be possible that these ``extreme''
configurations are unstable (tackling the problem of stability is
indeed one of the points in which the results of
Ref.~\cite{BH_plus_ring} may be improved in the future), but we do not
think this should be a major concern at this stage. Our viewpoint is
that considering such extreme configurations will provide a
testbed to investigate the practical problems of spacetime-mapping
through EMRI-gravitational waves. In particular, these configurations
will also help to understand better the confusion problem pointed out
by Glampedakis and Babak. As already stressed, while in quasi-Kerr
spacetimes this confusion disappears when dropping the time cut-off
and including radiation reaction~\cite{quasi_kerr_cutler}, in our case
it may still be present due to the practical difficulties of computing
radiation reaction in our spacetimes, which \textit{force} us to introduce
a cut-off in time.

We will see, however, that this confusion in the orbital parameters
appears in our spacetimes only for (equatorial) orbits far from the
black hole-torus system, whereas it disappears in the strong field
region. Nevertheless, we find another confusion problem, potentially
more worrisome as it involves the parameters of the black hole. Of
course, if we could replace the semi-relativistic approximation with a
rigorous solution of the linearized Einstein equations and a proper
treatment of self-force or radiation reaction, this confusion problem may
disappear as well. However, such a rigorous treatment is very hard to obtain
in generic stationary and axisymmetric spacetimes (see
Sect.~\ref{sec:rigorous_treatment}) and, as far as the self-force is
considered, even in pure Kerr.

This paper is organized as follows. In
Sect.~\ref{sec:rigorous_treatment} we show what the rigorous
treatment of EMRIs in non-vacuum, stationary and axisymmetric
spacetimes would be, and explain why this treatment has proved so hard that
nobody has pursued it so far. In Sect.~\ref{sec:background} we review
the non-Kerr spacetimes in which the problem of EMRIs has been considered
to date, ranging from approximate (Sect.~\ref{approximate_spacetimes})
to exact (Sect.~\ref{exact_spacetimes}) solutions of the Einstein
equations, and we introduce the non-pure Kerr spacetimes we will
use instead (Sect.~\ref{num_gen}).  In
Sect.~\ref{sec:semi-relativistic} we review the semi-relativistic
formalism used in Ref.~\cite{quasi_kerr} to compute gravitational
waves and explain how we adapted it to our purposes: in particular we
show how we integrated the geodesic equations and calculated kludge
waveforms, and (in Sect.~\ref{sec:overlap}) explain what the overlap
function and the dephasing time are. In Sect.~\ref{sec:comparison} we
explain in detail how we perform a comparison between our non-pure Kerr
spacetimes and pure Kerr spacetimes. A summary of our results with a
concluding discussion and the prospects of future work is presented in
Sect.~\ref{sec:res1} and~\ref{sec:res2}. Finally, in the Appendix we
review the connection between kludge waveforms and the linearized
Einstein equations.

Throughout this paper, we will use a system of units in which $G=c=1$. We will denote spacetime indices with Greek letters and space indices with Latin letters.

%-----------------------------------------------
\section{Waveforms from EMRIs in non-vacuum spacetimes}
\label{sec:rigorous_treatment}
%-----------------------------------------------

Let us consider a curved, non-vacuum spacetime with metric
$\boldsymbol g$ and with a characteristic lengthscale $M$ (for a
spacetime containing an SMBH, this scale clearly coincides with the
black hole mass). The spacetime is intrinsically not a vacuum one
because it contains a fluid with a
stress-energy tensor $\boldsymbol T^{\rm fluid}$. In addition,
consider the presence of a small body, such as a black hole with mass $m
\ll M$.\footnote{Note that in this context a small black hole can be
  treated as a small body despite being a singularity of
  spacetime~\cite{mi_sa_ta}.}  The small body will of course perturb the
geometry of spacetime: the metric $\boldsymbol {\widetilde g}$ of the
physical spacetime can therefore be written as the background metric
$\boldsymbol g$ plus some perturbations $\one\boldsymbol h$ of order
${O}\,(m/M)$, $\two\boldsymbol h$ of order ${O}((m/M)^2)$,
etc.:
\eq 
{\widetilde g}_{\mu\nu}=g_{\mu\nu} +\one h_{\mu\nu} + \two h_{\mu\nu}+
{O}((m/M)^3)\;.
\eeq 
Similarly, the small body will excite perturbations in the background
fluid: the perturbed stress-energy tensor of the fluid can be written
as
\eq
{\widetilde T}^{\rm fluid}_{\mu\nu}=T^{\rm fluid}_{\mu\nu} +\one
\delta T^{\rm fluid}_{\mu\nu} + \two\delta T^{\rm fluid}_{\mu\nu}+
       {O}((m/M)^3)\;.
\eeq
In what follows, the background metric $\boldsymbol g$ is used to
raise and lower tensor indices. For the sake of simplicity, we will also
drop the subscript $\one$ indicating first order quantities: in other
words, $h_{\mu\nu}\equiv \one h_{\mu\nu}$ and $\delta T^{\rm
  fluid}_{\mu\nu}\equiv \one\delta T^{\rm fluid}_{\mu\nu}$.

It is well-known that the stress-energy tensor of a
small body with mass $m$ following a trajectory $z^\mu(\widetilde{\tau})$ is
given by (see for instance Ref.~\cite{poisson})
\begin{multline}\label{eq:stress_en_part}
{\widetilde T}_{\rm small\,body}^{\alpha\beta}(x) =\\ m
\int {\widetilde P}^\alpha_{\ \mu}(x,z) {\widetilde P}^\beta_{\ \nu}(x,z)
{\widetilde u}^\mu {\widetilde u}^\nu\, \frac{\delta^{(4)}(x-z)}{(-\widetilde
g)^{1/2}}\, d\widetilde{\tau}
\end{multline}
where ${\widetilde P}^\alpha_{\ \mu}(x,z)$, $\widetilde{\tau}$ and ${\widetilde
u}^\mu\equiv d z^{\mu}/d\widetilde{\tau} $ are
respectively the parallel propagator from $z^\mu$ to $x^\mu$, the
proper time and the 4-velocity in the physical (i.e. perturbed)
spacetime. This stress-energy tensor can then be expanded in a
series in $m/M$: 
\eq
{\widetilde T}_{\rm small\,body}^{\alpha\beta}=T_{\rm small\,body}^{\alpha\beta}+
{O}\,((m/{M})^2)\;,
\eeq 
\begin{multline}\label{eq:stress_en_part_lowest_ord}
 T_{\rm small\,body}^{\alpha\beta}(x)=\\ m \int
{P}^\alpha_{\ \mu}(x,z) {P}^\beta_{\ \nu}(x,z) {u}^\mu {u}^\nu\,
 \frac{\delta^{(4)}(x-z)}{(-g)^{1/2}}\, d\tau \;,\end{multline}
where ${P}^\beta_{\ \nu}(x,z)$, $\tau$ and
$u^\mu=\mbox{d}z^\mu/\mbox{d}\tau $ are the parallel propagator,
proper time and 4-velocity in the background.

If the small body interacts only gravitationally with the matter
contained in the spacetime, its stress-energy tensor is conserved in
the physical spacetime:
\eq
\widetilde{\nabla}_\beta{\widetilde T}_{\rm small\,body}^{\alpha\beta}=0 \;,
\eeq
($\widetilde{\nabla}$ is the covariant derivative in the physical
spacetime). This implies that the small body follows a geodesic of the
physical, perturbed spacetime (see for instance Ref.~\cite{poisson} for a
formal proof): expanding the geodesic equations in the physical
spacetime (${\widetilde u}^\nu\widetilde{\nabla}_\nu {\widetilde
  u}^\mu=0$) into a series, it is possible to obtain, to
first-order in $m/M$,
\begin{multline}
\frac{D u^\mu}{d\tau} =-\frac{1}{2} \bigl( g^{\mu\nu} +
u^\mu u^\nu \bigr) \bigl( 2 \nabla_\rho h_{\nu\lambda} -
\nabla_\nu h_{\lambda\rho} \bigr) u^\lambda
u^\rho\\+{O}\,((m/{M})^2),\label{eq:modified_geod}
\end{multline} 
where $\nabla$ and $D/d\tau$ are the covariant derivative and the
total covariant derivative in the background.

Clearly, to zeroth order Eq.~\eqref{eq:modified_geod} reduces to the
geodesic equations in the background spacetime, but it deviates from
them at first-order. The right-hand-side of Eq.~\eqref{eq:modified_geod}
represents the so-called ``self-force'' and is physically due to
the interaction of the small body with its own gravitational field
$\boldsymbol h$; in the case of a small body orbiting around an SMBH,
this self-force is responsible for its inspiral towards the black
hole.

In order to compute the right-hand side of
Eq.~\eqref{eq:modified_geod} one needs to compute the metric
perturbation $\boldsymbol h$ and because this perturbation is produced
by the small body itself, some of its components will be divergent at the 
position of the small body. A regularization procedure to cure these divergences has been
derived~\cite{mi_sa_ta,quinn_wald} for the trace-reversed metric
perturbations
\begin{equation}
\label{rev_trace}
\bar{h}_{\mu\nu}\equiv
h_{\mu\nu}-\frac12\,h^\alpha_\alpha\,g_{\mu\nu}
\end{equation}
in the Lorenz gauge, which is defined as
\eq
\nabla_{\mu}\bar{h}^{\mu\nu}=0\;.
\eeq 
It should be noted that while this gauge allows one in principle to remove
the problem of divergences and has a number of other advantages (see
Ref.~\cite{barack} for an extensive list), self-force calculations are
extremely hard to perform in practice. Indeed, no general inspirals
have been computed so far using the regularized version of
Eq.~\eqref{eq:modified_geod}, not even in Schwarzschild or Kerr spacetimes
(see Ref.~\cite{poisson,self-force} for a review).  However, a simpler
approach can be followed in which only the dissipative part of the
self-force is taken into account, leading to the so-called ``adiabatic
approximation''~\cite{mino}\footnote{It should be
  noted that it is not yet clear whether the adiabatic approximation
  is accurate enough to compute waveforms for LISA as the conservative
  part of the self-force could have a secular effect as
  well~\cite{non_dissipative}.}. Within this approximation the small
body moves along a geodesic with slowly changing parameters (in Kerr,
these parameters are the energy $E$, the angular momentum $L_z$ and
Carter's constant $Q$). One of the advantages of the adiabatic
approximation is that it prescribes a way to compute the evolution of
these parameters, revealing that their changes $\dot E$ and $\dot L_z$
(with the dot being the derivative with respect to the coordinate time
$t$) correspond to the energy and angular momentum carried away by
gravitational waves~\cite{galtsov}. The change in Carter's constant
$\dot Q$, on the other hand, is harder to compute, although an
explicit formula has been recently derived~\cite{Qdot}.
 
The first-order metric perturbation $\boldsymbol h$ can be
computed as a solution of the linearized Einstein
equations~\cite{poisson2}
\begin{multline}\label{eq:box_h}
\Box\, \bar{h}^{\alpha\beta} + 2 R_{\mu\ \nu}^{\ \alpha\ \beta}
	\bar{h}^{\mu\nu} + S_{\mu\ \nu}^{\ \alpha\ \beta}
        \bar{h}^{\mu\nu} = \\
	-16\pi (\delta T_{\rm fluid}^{\alpha\beta}+T_{\rm
          small\,body}^{\alpha\beta})\;,
\end{multline}
where
\begin{gather}
S_{\mu\alpha\nu\beta} = 2 G_{\mu(\alpha} g_{\beta)\nu} -
R_{\mu\nu} g_{\alpha\beta}  - 2 g_{\mu\nu} G_{\alpha\beta}\;,\\
\Box\equiv g^{\mu\nu}\nabla_\mu\nabla_\nu%\;.
\end{gather}
($R_{\mu\nu\alpha\beta}$, $R_{\mu\nu}$ and $G_{\mu\nu}$ are the
background Riemann, Ricci and Einstein tensors).  Note that self-force
effects are not contained in~\eqref{eq:box_h}, which is a first order equation.
In fact, since the stress-energy tensor of the small
body at the lowest order, $\boldsymbol T_{\rm small\,body}$, is an
intrinsically first-order quantity [remember the factor $m$ appearing
in Eq.~\eqref{eq:stress_en_part_lowest_ord}], the small body's
contribution can be computed using a zeroth-order expression of
$u^\mu$ or, equivalently, by solving the geodesic equations for the
background metric. In addition to the calculation of the small body's
contribution, a consistent solution at first-order for the EMRI
problem in a curved and non-vacuum spacetime requires the solution of
the fluid perturbation $\boldsymbol \delta \boldsymbol T_{\rm
  fluid}$. This can be computed by imposing the conservation of the
stress-energy tensor of the fluid, $\widetilde{\nabla}_\beta{\widetilde T}_{\rm
  fluid}^{\alpha\beta}=0$, which gives, to first order,
\begin{multline} -16\pi\, { \nabla_\beta\,\delta
    T^{\alpha\beta}_{\rm
      fluid}}=2G^{\beta\sigma}\nabla_\sigma\bar{h}_{\beta}^\alpha\\-2
  G^{\alpha\beta}\partial_\beta\bar{h}-
	R^{\beta\sigma}\nabla_\gamma\bar{h}_{\beta\sigma}g^{\gamma\alpha}\;.
\label{eq:euler}
\end{multline}
It is not difficult to realize, using Eqs.~\eqref{eq:box_h} and
\eqref{eq:euler}, that the Lorenz gauge condition is conserved since it
satisfies a homogeneous equation
\eq\Box(\nabla_\beta\bar{h}^{\alpha\beta})+
R^\alpha_\mu\nabla_\beta\bar{h}^{\mu\beta}=0\;.
\eeq
To summarize, the solution of Eqs.~\eqref{eq:box_h} with the
right-hand-side given by Eqs.~\eqref{eq:euler} and the zeroth-order
contribution of Eq.~\eqref{eq:modified_geod}, provides the complete
and consistent solution of the EMRI problem at first-order in
$m/M$. Unfortunately, for situations of practical interest, such as for
the observations of EMRIs performed by LISA, these first-order waveforms
would be sufficiently accurate only for a few days or
weeks~\cite{quasi_kerr,T_RR_drasco}, imposing, at least in principle,
the need for the solution of second-order equations.

Clearly, the solution of the second-order perturbation equations is
much harder to obtain as these will have a schematic generic form of
the type
\eq 
{\cal D}[\two {\boldsymbol h}]= {O}
	\left(\nabla{\boldsymbol h}\nabla{\boldsymbol h},  
       {\boldsymbol h}\nabla\nabla{\boldsymbol h}\right)\;,
\eeq
where ${\cal D}[\two{\boldsymbol h}]$ is a differential operator acting
on $\two {\boldsymbol h}$.

One could na\"ively try to solve this equation by imposing a gauge
condition on $\two \boldsymbol h$ and using the Green function of the
$\cal D$ operator, but the formal solution obtained in this way would
be divergent at every point because of the divergences of the first-order
perturbation $\boldsymbol h$ at the small body's position. A
regularization procedure to cure these divergences is
known~\cite{2nd_order}, but it has not yet been applied in practical
calculations.

An alternative to the solution of the full second-order perturbation
equations entails introducing the deviations from geodetic
motion in the right-hand-side of Eq.~\eqref{eq:box_h}. This approach
is clearly not consistent, but hopefully accurate enough if the ratio
$m/M$ and consequently the deviations from geodetic motion
are sufficiently small. This is indeed what was done by Drasco
and Hughes~\cite{hughes_drasco}, who used the adiabatic
approximation and a simplified formula for $\dot Q$ to compute the
deviations from geodetic motion, inserting them in the right-hand-side
of the Teukolsky equation~\cite{teukolsky_formalism} and then solving for
\textit{first-order} perturbations.

While very appealing, as it provides a simple way to improve upon a
purely first-order calculation, we will not follow this approach
here. Rather, we will perform our calculations within a
semirelativistic (``kludge'') approximation to Eq.~\eqref{eq:box_h}, using
however as a  background spacetime a non-trivial departure from a
pure-Kerr solution. The properties of this spacetime and of
alternative formulations of non-Kerr spacetimes will be discussed
in detail in Sect.~\ref{sec:background}, while a brief description of
our semi-relativistic approach will be presented in
Sect.~\ref{sec:semi-relativistic}.

%-----------------------------------------------
\section{Modelling the background spacetime}
\label{sec:background} 
%-----------------------------------------------

The discussion made in the previous section assumes that a background
spacetime ${\boldsymbol g}$ is known and this is traditionally
assumed to be a ``pure-Kerr'' solution. However, this is not the only
possibility. Indeed, in order to investigate LISA's ability to
detect deviations from Kerr, a number of attempts have been made
recently to replace the Kerr metric with other stationary solutions
representing reasonable deviations from a single rotating
black hole in vacuum. In what follows we will briefly review  these attempts
and discuss a novel one based on the use of highly-accurate numerical
solutions of the Einstein equations for spacetimes containing a black
hole and a compact torus (see Sect.~\ref{num_gen}).

%===============================================
\subsection{Approximate non-Kerr spacetimes}
\label{approximate_spacetimes}
%===============================================

One first attempt to go beyond a pure-Kerr model for the central
massive object was suggested by
Ryan~\cite{ryan_multipoles,ryan_3,ryan_4}, who considered a
general stationary, axisymmetric, asymptotically flat, vacuum
spacetime, which can be used to describe the gravitational field around a central distribution of matter, and its expansion in terms of the mass multipoles $M_\ell$
and of the current multipoles $S_\ell$~\cite{multipoles}. If one assumes reflection symmetry,
the odd $M$-moments and even $S$-moments are identically
zero \cite{Kordas,Meinel}, so that the non-vanishing moments are the mass $M_0=M$, the mass
quadrupole $M_2$ and the higher-order even multipoles $M_4, M_6,
\ldots$, as well as the angular momentum $S_1=J$, the current octupole
$S_3$ and the higher-order odd multipoles $S_5, S_7, \ldots$. The
metric can then be written as
\begin{multline}
ds^2 = -e^{\gamma + \delta}\,dt^2 + e^{2\alpha}\,( dr^2 + r^2\,d\theta^2)
\\+ e^{\gamma -\delta}\,r^2\sin\theta^2( d\phi -\omega dt)^2\;,
\label{eq:metric_multi}
\end{multline}
where the potentials $\gamma,\delta,\omega,\alpha $ depend only on $r$
and $\theta$. Each of them can expanded in terms of the multipole
moments: for example
\begin{equation}
\delta = \sum_{n=0}^{+\infty}\,
-2\frac{M_{2n}}{r^{2n+1}}\,P_{2n}(\cos\theta)+(\mbox{higher order
  terms})\;,
\label{eq:metric_exp1}
\end{equation}
\begin{multline}
\omega = \sum_{n=1}^{+\infty}\,
-\frac{2}{2n-1}\,\frac{S_{2n-1}}{r^{2n+1}}\,\frac{P_{2n-1}^1(\cos\theta)}
{\sin\theta}\\+ (\mbox{higher order terms})\;,
\label{eq:metric_exp2}
\end{multline}
where $P_{2n},~P^{1}_{2n-1}$ are the Legendre and the associated
Legendre polynomials and where only the lowest-order $1/r$-dependence of
each multipole moment is shown. 

The multipoles are related to the interior matter distribution and could in principle be computed by solving the Einstein equations.
In the particular case of a Kerr spacetime, all the multipole moments are
trivially related to the first two, mass and angular momentum, by the
following relation:
\eq
M_\ell+{\rm i}S_\ell=M\left({\rm i}\frac JM\right)^\ell\;.
\eeq 
This is the celebrated ``no hair'' theorem: an (uncharged) stationary
black hole is uniquely determined by its mass and spin. Deviations from the
Kerr metric can be therefore detected by measuring the mass, spin and
higher order moments of the central massive object.

While general and very elegant, this approach has serious drawbacks in the
strong-field region near the central massive object, which is clearly
the most interesting one. In fact, this is the region which will be
mapped by LISA and where the spacetime could be significantly
different from Kerr.  The origins of these drawbacks are rather
apparent when looking at
Eqs.~\eqref{eq:metric_multi}--\eqref{eq:metric_exp2}, which are in
practice an expansion in powers of $1/r$ around a Minkowski
spacetime. As a result, an accurate representation of the strong field
region necessarily requires the inclusion of many multipoles.

Another approach to the modelling of a non-Kerr background
spacetime was recently suggested by Babak and Glampedakis in
Ref.~\cite{quasi_kerr}, and is based on the use of the Hartle-Thorne
metric~\cite{HT_metric}. This metric describes the spacetime outside
slowly rotating stars, includes as a special case the Kerr metric at order
${O}(a^2)$, where $a\equiv J/M^2$, and is accurate up to the mass
quadrupole moment. In order to isolate the quadrupolar deviation with
respect to Kerr, the Hartle-Thorne metric can be rewritten in terms of
the parameter $\epsilon$ defined as
\eq
\label{eq:eps} 
Q = Q^{\rm Kerr} - \epsilon\,M^3\;,\qquad Q^{\rm Kerr}= -\frac{J^2}{M}
\eeq
where $M$, $J$ and $Q\equiv M_2$ are the mass, the angular momentum
and the mass quadrupole moment, respectively. Since for Kerr $Q=Q^{\rm Kerr}$,
$\epsilon$ can be used as a lowest-order measure of the deviation of
the spacetime from a Kerr solution. The metric expressed in this
way can be further rewritten in ``quasi-Boyer-Lindquist
coordinates'', i.e. coordinates reducing to
Boyer-Lindquist coordinates if $\epsilon=0$. This procedure then leads
to the ``quasi-Kerr'' metric
\begin{multline}
\label{eq:quasi_kerr}
g^{\rm quasi-Kerr}_{\mu\nu}=g^{\rm Kerr}_{\mu\nu}+\epsilon\,
h_{\mu\nu}\\+{O}(a\,\epsilon,\,\epsilon^2) +
	{O} (\delta M_{\ell\geq4}, \delta S_{\ell\geq3})\;,
\end{multline}
where $g^{\rm Kerr}_{\mu\nu}$ is the Kerr metric in Boyer-Lindquist
coordinates, $\epsilon h_{\mu\nu}$ is the deviation from it and $\delta
M_{\ell\geq4}$, $\delta S_{\ell\geq3}$ are the deviations of the
higher-order multipoles from those of a Kerr spacetime. Stated
differently, the quasi-Kerr metric consists of a Kerr solution plus a
small difference in the mass quadrupole expressed by the parameter
$\epsilon$, while neglecting any deviations from Kerr in the
higher-order multipoles $M_4, M_6, \ldots$, and $S_3, S_5, \ldots$,
etc.

Because this approach does not involve any expansion in powers of
$1/r$, it can be used in the strong-field regions as long as the
central massive object is slowly rotating. Furthermore, it has the
great advantage of being straightforward to implement, leaving the
mass quadrupole parameter $\epsilon$ as the only adjustable
one. However, it has the drawback that it does not include any
deviations in the multipoles higher than the quadrupole with respect
to the multipoles of pure Kerr, which could be important in the strong
field regime.

%===============================================
\subsection{Exact non-Kerr spacetimes}
\label{exact_spacetimes}
%===============================================

A different approach to the modelling of the background consists
instead of using \textit{exact} solutions of the Einstein equations:
these spacetimes of course behave well in the strong field regime,
since they are not based on any series expansions.

Few attempts have been made in this direction. However, Kesden,
Gair \& Kamionkowski~\cite{gair_boson_star} considered spacetimes
containing non-rotating boson stars and found that the gravitational
waves produced by EMRIs look qualitatively different from the pure
black hole case.  The spherical boson stars they consider 
are in fact identical to Schwarzschild spacetimes outside their surfaces,
making them
indistinguishable from black holes during the initial stages of an
EMRI. However, for a black hole the event horizon
prevents any observations of the inspirals after the final
plunge. Because boson stars are horizonless however, many orbits inside the
interior are expected if the small body interacts only gravitationally
with the scalar field out of which the star is made: the ``smoking
gun'' for a boson star would therefore be that gravitational waves
from the inspiral are observed after the plunge. Gravitational waves from
such an event could not be interpreted as an inspiral into a black
hole with different parameters, because the first part of the
inspiral is identical to the usual black-hole inspiral.

Another attempt was made by Collins and Hughes in
Ref.~\cite{bumpyBHs}.  The analytical ``bumpy black holes'' they build
are objects that are \textit{almost} Schwarzschild black holes, but
have some multipoles with a `wrong' value. These spacetimes reduce to
the usual Schwarzschild black hole spacetimes in a natural way, by
sending the ``bumpiness'' of the black hole to zero, but unfortunately
they require naked singularities: although they are not expected to
exist in nature, ``bumpy black holes'' could be useful as
``straw-men'' to set up null experiments and test deviations from pure
Kerr using EMRIs.

%===============================================
\subsection{A self-gravitating torus around a rotating black hole}
\label{num_gen}
%===============================================

A different and novel approach to the modelling of a background,
non-Kerr spacetime is also possible and will be the one adopted in
this paper. In particular, we exploit the consistent numerical
solution of the full Einstein equations describing a spacetime with an
axisymmetric black hole and a compact, self-gravitating torus of
comparable mass and spin. These solutions have been recently obtained
to great accuracy with a numerical code using spectral methods.
In general, the numerical solution will produce a solution
of the Einstein equations representing an axisymmetric and stationary
spacetime containing a rotating black hole and a constant-density,
uniformly rotating torus of adjustable mass and spin. The metric of
this \textit{non-pure Kerr spacetime} in quasi-isotropic (QI)
coordinates is generically given by
\begin{multline}
ds^2 = -e^{2\nu}dt^2 +
{r_{_{\rm QI}}}^2 \sin^2\theta B^2 e^{-2\nu} \left(d\phi -
\omega dt\right)^2\\ + e^{2\mu}\left(d{r\mbox{\tiny
    $_{QI}$}}^2+{r_{_{\rm QI}}}^2d\theta^2
\right)\;,
\label{eq:metric}
\end{multline} 
where $\nu$, $\mu$, $B$ and $\omega$ are functions of the radial
quasi-isotropic coordinate $r_{_{\rm QI}}$ and $\theta$.
The procedure for obtaining such numerical solutions is described in
detail in Ref.~\cite{BH_plus_ring} and we here provide only a summary
of the main ideas.

The entire spacetime outside of the horizon is described by a metric
in Weyl-Lewis-Papapetrou coordinates as in Eq.~\eqref{eq:metric} or~\eqref{eq:metric_multi}. We
fix our coordinates uniquely by stipulating that the first derivatives
of the metric functions be continuous at the ring's surface and by
choosing the location of the horizon to be a coordinate sphere
$r_{\text{\tiny QI}}=\text{constant}\equiv r_{_{+,\rm QI}}$.
Specifying the boundary conditions $B=0$, $e^{2\nu}=0$ and
$\omega=\text{constant}$ on this sphere ensures that it is indeed a
black hole horizon. We further assume reflection symmetry with
respect to the equatorial plane.

We are interested only in the metric outside of the black hole and
determine it using a multi-domain spectral method.  One of the domains
coincides precisely with the interior of the homogeneous, uniformly
rotating perfect fluid ring, the boundary of which must be solved for as part of the
global problem. This choice is important in order to avoid Gibbs
phenomena. The vacuum region (outside the horizon) is divided up into
four subdomains with three fixed boundaries separating them. This
somewhat arbitrary choice enables us to resolve functions more
accurately in the vicinities of both the ring and the black hole
according to the scale determined by each object itself. One of the
four vacuum domains extends out to infinity and is then
compactified. A mapping of each domain onto a square is chosen in such
a way as to avoid steep gradients in the functions being solved for.

The Einstein equations together with the specification of asymptotic
flatness and the aforementioned boundary conditions provide us with a
complete set of equations to be solved for. The metric functions and
the function describing the ring's boundary are expanded in terms of a
finite number of Chebyshev polynomials. By specifying physical
parameters to describe a configuration and demanding that the
equations be fulfilled at collocation points on these five domains, we
get a non-linear system of algebraic equations determining the
coefficients in the expansion of the functions. We solve this system
using a Newton-Raphson method where an existent neighbouring solution
provides the initial guess (see Ref.~\cite{BH_plus_ring} for more details
and for a discussion of how to obtain the first ``initial guess'').

Note that throughout this paper, the masses and angular momenta of the
black hole, $M_{_{\rm BH}}$ and $J_{_{\rm BH}}$, of the torus, $M_{_{\rm
    Torus}}$ and $J_{_{\rm Torus}}$, and of the whole system, $M_{_{\rm
    tot}} \equiv M_{_{\rm BH}}+M_{_{\rm Torus}}$ and $J_{_{\rm tot}}
\equiv J_{_{\rm BH}}+J_{_{\rm Torus}}$, are the ``Komar'' masses and
angular momenta~\cite{komar,bardeen}. We note that the definition of the
mass of a single object in General Relativity can be quite slippery,
especially when this measure is not an asymptotic one. We also recall
that while the \textit{total} Komar mass of the system coincides with the
familiar ``ADM'' mass~\cite{adm}, other definitions are possible for the
\textit{single} masses of the torus and the black hole. As an example, it
is possible to define the ``irreducible mass'' of the black hole as
$M_{_{\rm irr}} \equiv [A_{_{+}}/(16\pi)]^{1/2}$, where $A_{_{+}}$ is the
horizon's area~\cite{christodoulou}, and then define the total mass of
the black hole as $M_{_{\rm hole}} \equiv [M_{_{\rm irr}}^2 + (J_{_{\rm
    BH}}/ (2 M_{_{\rm irr}}))^2]^{1/2}$ (Note that this latter definition
coincides with the Komar mass for an isolated Kerr black hole.).
Similarly, the mass of the torus can also be measured in terms of the
``baryonic'' mass $M_{_{\rm baryon}}=\int\rho\, u^t\sqrt{-g}\,d^3x$
($\rho$ being the baryonic mass density and $\boldsymbol u$ the
4-velocity of the fluid of the torus). This mass is simply a measure of
the number of baryons, it does not include the gravitational binding of
the object, and thus can also be rather different from the corresponding
Komar mass.

Overall, we have found that even in our non-pure Kerr spacetimes, all the
definitions of the mass of the black hole agree rather well. In
particular, in the spacetimes $A$ and $B$ we will consider in
sections~\ref{sec:res1}-\ref{sec:res2} (cf.
table~\ref{table:spacetimes_data}) we have $M_{_{\rm BH}}=0.413$,
$M_{_{\rm irr}}=0.457$, $M_{_{\rm hole}}=0.468$ and $M_{_{\rm BH}}=0.1$,
$M_{_{\rm irr}}=0.1007$, $M_{_{\rm hole}}=0.1007$, respectively. On the
other hand, the Komar mass and the baryonic mass of the torus have been
found to be different with $M_{_{\rm Torus}}=0.121$ and $M_{_{\rm
    baryon}}=0.0578$ in spacetime $A$ and $M_{_{\rm Torus}}=0.007$ and
$M_{_{\rm baryon}}=0.00656$ in spacetime $B$. As mentioned above, there
is no reason to expect these two measures to be close and it is only
interesting that this happens under certain circumstances (as in
spacetime $B$, for example). Furthermore, these differences are not going
to affect our analysis, which will never use the single mass of the
torus.

%-----------------------------------------------
\section{The semi-relativistic approach}
\label{sec:semi-relativistic} 
%-----------------------------------------------

Although the procedure outlined in 
Sect.~\ref{sec:rigorous_treatment} to calculate the waveforms from an
EMRI in a non-vacuum spacetime is the only mathematically correct one, it
has never been applied in practice, not even to first order. Such an
approach, in fact, involves the solution of a complicated system of 14
coupled partial differential equations [Eqs.~\eqref{eq:box_h} and
  \eqref{eq:euler}] and while this can in principle be solved,
alternative solutions have been traditionally sought in the
literature. A very popular one is the ``semi-relativistic'' approach,
which leads to the so-called ``kludge''
waveforms~\cite{ruffini_and_sasaki, kostas_review, babak_kludge} and
which we will also adopt hereafter.

In essence, the semi-relativistic approach consists in considering
geodetic motion for the small body (including, when possible,
corrections to account at least approximately for the effects of
radiation reaction) and in calculating the emitted gravitational waves
as if the small body were moving in a Minkowski spacetime. This latter
assumption requires a mapping between the real spacetime and
the Minkowski spacetime: in the pure Kerr case, this is obtained by
identifying Boyer-Lindquist coordinates with the spherical coordinates
of the Minkowski spacetime. The waveforms
are then computed using the standard quadrupole, octupole or higher order
formulas\footnote{Note that comparisons with Teukolsky-based waveforms
  in Kerr show that the inclusion of multipoles higher than the
  octupole does not improve kludge
  waveforms~\cite{babak_kludge}. Because of this, contributions only
  up to the octupole are used here to calculate kludge
  waveforms.}. Waveforms obtained in this way are commonly referred to
as ``kludge'' waveforms~\cite{ruffini_and_sasaki, kostas_review,
  babak_kludge}

A justification of this procedure in terms of the Einstein equations
is given in the Appendix.  However, the strongest
motivation for introducing kludge waveforms is the surprising
agreement they show with the accurate waveforms that can be computed
in a Kerr spacetime using the Teukolsky
formalism~\cite{babak_kludge}. In view of this, kludge waveforms
represent the natural first approach to model the emission from EMRIs
in non-pure Kerr spacetimes and will be used throughout this work.

As mentioned in Sect.~\ref{sec:rigorous_treatment}, the adiabatic
approximation offers a simple way to include radiation-reaction
effects in a Kerr spacetime. More specifically, if we denote Kerr
geodesics by $x_{\rm geod}^\mu(t,E,L_z,Q)$, the trajectory of the
small body is then corrected to be $x^\mu(t)=x_{\rm
  geod}^\mu(t,E(t),L_z(t),Q(t))$, that is, a geodesic with varying
parameters. The accurate calculation of the fluxes $\dot E$, $\dot{L} _z$ and
$\dot Q$ is rather involved~\cite{Qdot,hughes_drasco}, but approximate
ways to compute them have also been
suggested~\cite{ryan_fluxes,kludge_first_paper,GG_kludge_fluxes}. Although Barack
\& Cutler~\cite{quasi_kerr_cutler} have recently proposed including
radiation reaction in quasi-Kerr spacetimes by using post-Newtonian
fluxes in which the leading-order effect of the quadrupole of the
spacetime is taken into account, it is still unclear at this stage how
accurately the fluxes for a Kerr or a quasi-Kerr spacetime could
describe the non-geodetic motion of an EMRI around our black hole-torus
systems. Because of this, we have here preferred to consider the
simplest scenario and thus model the motion of the small body as a
pure geodesic with equations of motion that in the spacetime
Eq.~\eqref{eq:metric} are given by
\begin{eqnarray}
&&\hskip -0.75cm
\frac{dt}{d\tau}=-g^{tt}\tilde\epsilon+g^{t\phi}\tilde\ell\;,
\label{eq:geo_t}\\
&&\hskip -0.75cm
\frac{d\phi}{d\tau}=-g^{t\phi}\tilde\epsilon+g^{\phi\phi}\tilde\ell\;,
\label{eq:geo_phi}\\
&&\hskip -0.75cm
\frac{d^2r}{d\tau^2}=-\Gamma^r_{rr}\left(\frac{dr}{d\tau}\right)^2
	-\Gamma^r_{\theta\theta}\left(\frac{d\theta}{d\tau}\right)^2
        -2\Gamma^r_{r\theta}\frac{dr}{d\tau}\frac{d\theta}{d\tau}\nonumber\\
&&\hskip 0.25cm -\Gamma^r_{tt}\left(\frac{dt}{d\tau}\right)^2-
        \Gamma^r_{\phi\phi}\left(\frac{d\phi}{d\tau}\right)^2
        -2\Gamma^r_{t\phi}\frac{dt}{d\tau}\frac{d\phi}{d\tau}\;,
\label{eq:geo_r}\\
&&\hskip -0.75cm
\frac{d^2\theta}{d\tau^2}=-\Gamma^\theta_{rr}\left(\frac{dr}{d\tau}\right)^2
 	-\Gamma^\theta_{\theta\theta}\left(\frac{d\theta}{d\tau}\right)^2
         -2\Gamma^\theta_{r\theta}\frac{dr}{d\tau}\frac{d\theta}{d\tau}\nonumber\\
&&\hskip 0.25cm    -\Gamma^\theta_{tt}\left(\frac{dt}{d\tau}\right)^2-
         \Gamma^\theta_{\phi\phi}\left(\frac{d\phi}{d\tau}\right)^2
         -2\Gamma^\theta_{t\phi}\frac{dt}{d\tau}\frac{d\phi}{d\tau}
 \label{eq:geo_theta}\;,
\end{eqnarray}
where $r\equiv{r_{_{\rm QI}}}$ is the radial quasi-isotropic
coordinate, $\tau$ is the proper time, the $\Gamma$'s are the
Christoffel symbols and $\tilde\epsilon$ and $\tilde\ell$ are the
energy and angular momentum per unit mass as measured by an observer
at infinity.

The resulting geodesics can be labelled with seven parameters: four
refer to the initial position $t_0,\phi_0,r_0,\theta_0$ and the
remaining three identify the initial 4-velocity. In the case, which we will consider in this paper, of bound stable orbits in the equatorial plane, only
five parameters would remain. However, because of stationarity and axisymmetry it is not restrictive to fix $t_0=\phi_0=0$ and $r_0=r_p$, being $r_p$ the periastron radius. Therefore, except for a sign to distinguish between prograde ($\dot\phi>0$) and
retrograde ($\dot\phi<0$) orbits, equatorial bound stable geodesics can be characterized by two parameters only, which we can choose to be the so-called 
``latus rectum'' $p_{_{\rm QI}}$ and the ``eccentricity'' $e_{_{\rm QI}}$, which
are related to the coordinate radii at apoastron and periastron by
$r_a\equiv p_{_{\rm QI}}/(1-e_{_{\rm QI}})$ and $r_p\equiv p_{_{\rm QI}}/(1+e_{_{\rm QI}})$.

Clearly, kludge waveforms computed from pure geodetic motion are
expected to be accurate only below the timescale over which
radiation-reaction effects become apparent and make our waveforms
differ significantly from the real signal. A simple way to estimate
this timescale exploits the
concept of ``overlap'' between two waveforms, which will be
introduced in Sect.~\ref{sec:overlap}.

An important comment needed here is instead on the coordinate mapping
used in calculating kludge waveforms. As already mentioned, this
mapping has a straightforward realization in a Kerr spacetime, where
the BL coordinates are associated with the spherical coordinates of a
Minkowski spacetime. In a similar manner, in our non-pure Kerr
spacetimes we transform the solution of the geodesic equations from QI
coordinates to ``quasi-Boyer-Lindquist'' (QBL) coordinates,
i.e.  coordinates that reduce to BL coordinates in the
absence of the torus. These coordinates are then identified with the
spherical coordinates of a Minkowski spacetime as in
Ref.~\cite{quasi_kerr} and used to compute kludge waveforms.

\begin{figure}
\begin{center}
\vspace{1.0 cm}
\includegraphics[width=8.25 cm]{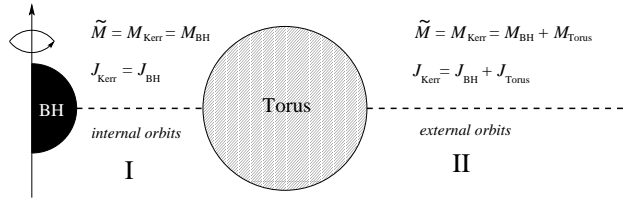}
\vspace{1.0 cm}
\caption{Schematic classification of the two regions of the
  spacetime. For equatorial orbits in region I (i.e. internal orbits) the mass
  and angular momentum of the Kerr black hole coincide with the mass
  and angular momentum of the black hole. For equatorial orbits in region II
  (i.e. external orbits) the mass and angular momentum of the Kerr
  black hole coincide with the total mass and angular momentum of the
  black hole-torus system.
\label{spacetime_regions}}
\end{center}
\end{figure}

Fortunately, the transformation from QI to QBL coordinates is
straightforward and involves only a change in the radial coordinate:
\eq
\label{eq:QI2QBL} r_{_{\rm QBL}}=
	r_{_{\rm QI}}+\widetilde{M}+
        \frac{r^2_{_{+,\rm QI}}}{r_{_{\rm QI}}}\;, 
\eeq
where 
$\widetilde{M}$ is a parameter that reduces to the mass of the central
black hole in the absence of the torus. Clearly, this mapping suffers
from an intrinsic ambiguity as the mass $\widetilde{M}$ could be either
associated with the mass of the black hole or with the total mass of the
system, or even with a combination of the two. Although all the
choices are essentially equivalent when the torus is very light, this
is not necessarily the case for some of the configurations considered
here, for which the torus has a mass comparable with that of the black
hole. Since the parameter $\widetilde{M}$ is, at least in a Newtonian
sense, the gravitational mass experienced by the small body, we have
here followed a pragmatic approach and set $\widetilde{M}=M_{_{\rm
    tot}}$ for equatorial  orbits with
periastron larger than the outer edge of the torus, which we will
refer to as the \textit{``external orbits''}. Conversely, we have set
$\widetilde{M}=M_{_{\rm BH}}$ for what we will refer to as the
\textit{``internal orbits''}, that is equatorial  orbits with both periastron and
apoastron between the inner edge of the torus and the horizon. This
classification is summarized schematically in
Fig.~\ref{spacetime_regions}, which shows the two regions into which
the spacetime has been divided and the corresponding values of
$\widetilde{M}$. This choice is clearly an operative ansatz, but we
have checked to see that its influence on our results is indeed negligible
and a detailed discussion of this will be presented in
Sect.~\ref{sec:comparison}.

Finally, we note that we have not considered orbits crossing the torus
because the non-gravitational interaction between the small body and the
fluid would cause deviations from geodetic motion which are not easy
to model.

%===============================================
\subsection{Overlap and dephasing time}
\label{sec:overlap}
%===============================================

In order to compare (kludge) waveforms computed in non-pure Kerr
spacetimes with (kludge) waveforms computed in Kerr spacetimes, we
follow the procedure proposed in Ref.~\cite{quasi_kerr} and make use
of the so-called \textit{overlap function}. Its meaning can be
best understood through the more familiar concept of SNR, which we will now briefly review.

We recall that if a signal $s(t)$ is the sum of a gravitational wave
$h(t)$ and of some Gaussian noise $n(t)$, the SNR for a template $\hat
h(t)$ is given by~\cite{cutler}
\eq 
\frac{S}{N}[\hat h] = \frac{
  \int \hat h(t)\,w(t-\tau)\,s(\tau)\, d\tau dt}{{\rm rms}\left[ \int
  \hat h(t) \, w(t-\tau)\, n(\tau) \, d\tau dt\right]}=\frac{(\hat
  h,s)}{(\hat h,\hat h)^{1/2}}\label{eq:S2N} 
\eeq 
where $w(t)$ is Wiener's optimal filter (i.e. the Fourier transform of
the function $w(t)$ is given by $\tilde{w}(f)=1/S_n(f)$, with $S_n(f)$
being the spectral sensitivity of the detector), ``rms'' denotes the
root mean square and the internal product
``$(\phantom{a},\phantom{a})$'' can be defined in terms of the Fourier
transforms (which are denoted by a ``tilde''):
\eq\left( h_1 , h_2 \right) \equiv 2 \int_0^{\infty} \frac{{\tilde
    h}_1^*(f) {\tilde h}_2(f) + {\tilde h}_1(f) {\tilde
    h}_2^*(f)}{S_n(f)} 
df\;.
\eeq 

Clearly, the SNR of Eq.~\eqref{eq:S2N} is a Gaussian random variable
with zero average and unit variance if no gravitational wave signal is
present. On the other hand, in the presence of a gravitational wave
the expected value for the SNR is nonzero with a time average given by
\eq\left\langle
\frac{S}{N}[\hat h]\right\rangle=
\frac{(\hat h,h)} {(\hat h,\hat h)^{1/2}} + \frac{(\hat h,n)} {(\hat
  h,\hat h)^{1/2}}  = \frac{(\hat h,h)} {(\hat h,\hat h)^{1/2}} \;. 
\eeq 
If $\alpha$ measures the SNR for a template $\hat h(t)$ ``matching''
the gravitational wave $h(t)$ perfectly, i.e. $\langle
S/N\rangle =(h,h)^{1/2}\equiv \alpha$, any ``mismatch'' between $\hat
h(t)$ and $h(t)$ will degrade the SNR ratio to \mbox{$\langle
  S/N\rangle = \alpha\, {\cal O}(h,\hat h)$}, where the overlap
function $\cal O$ is defined as
\eq\label{eq:overlap} {\cal O}(h,\hat h)\equiv 
\frac{(h,\hat h)}{(\hat h,\hat h)^{1/2}(h, h)^{1/2}}\;.  
\eeq

The same logic can now be used to quantify the differences between
kludge waveforms computed in different spacetimes. More specifically,
if we label with ``1'' a waveform computed in a non-pure Kerr  
spacetime and with ``2'' the closest equivalent in a Kerr spacetime,
the overlap between the two ${\cal O}\,(h_1,h_2)\equiv
{(h_1,h_2})/[(h_1, h_2)^{1/2}(h_1, h_2)^{1/2}]$ will express how much
SNR is lost by an observer match-filtering a black hole-torus signal with
a pure Kerr template. Stated differently, ${\cal O}(h_1,h_2)=1$ if the
two waveforms are identical, while ${\cal O}(h_1,h_2)=0$ if they are
totally uncorrelated and ${\cal O}(h_1,h_2)=-1$ if they are perfectly
anticorrelated.

Having introduced the concept of overlap function, we can proceed
to an operative definition of the timescale below which kludge
waveforms computed from pure geodetic motion are expected to be
accurate. This timescale, usually referred to as the ``dephasing
time'' $\tau_{\rm d}$, is defined as the time at which the overlap
between two waveforms in the Kerr spacetime, one computed considering
geodetic motion and the other one including radiation reaction
effects, drops below 0.95 (this is indeed the threshold used to build
template banks~\cite{minimal_match}). Clearly, the dephasing time will
be different for external and internal orbits and also in this case
attention needs to be paid to the mappings between non-pure and pure Kerr spacetimes. Following the same logic
discussed in the previous section, we calculate $\tau_{\rm d}$ for an
external equatorial orbit 
in our non-pure Kerr spacetime by considering the equatorial orbit
with the same latus rectum and eccentricity in the Kerr spacetime\footnote{The latus rectum and the eccentricity
  are assumed to be in BL coordinates in pure Kerr and in QBL in non-pure Kerr spacetimes.} with mass $M_{_{\rm Kerr}} = M_{_{\rm tot}}$
and spin $J_{_{\rm Kerr}} = J_{_{\rm tot}}$.
 On the other hand, for an internal orbit we calculate
$\tau_{\rm d}$ by considering the orbit with the same latus rectum and eccentricity in
the Kerr spacetime with mass $M_{_{\rm Kerr}}=M_{_{\rm BH}}$ and spin
$J_{_{\rm Kerr}} = J_{_{\rm BH}}$. As we will explain, in this case we have also
looked into the influence that this association has on the overall results
presented in Sect.~\ref{sec:comparison}.

In order to compute the dephasing time, we used the approximate Kerr
fluxes proposed in Ref.~\cite{GG_kludge_fluxes}, which are based on
post-Newtonian expansions and fits to fluxes computed rigorously with
the Teukolsky formalism.

%-----------------------------------------------
\section{Comparing pure and non-pure Kerr spacetimes}
\label{sec:comparison}
%-----------------------------------------------

The set of tools introduced in the previous sections, namely: the
kludge waveforms, the numerical solution of the Einstein equations for
spacetimes containing a black hole and a torus, and the overlap
function, can now be applied to determine to what extent LISA can
detect a difference between a pure and a non-pure Kerr spacetime.

Hereafter we will restrict our attention to equatorial, bound and
stable orbits, choosing the values of the mass and angular momentum
of the pure Kerr spacetime using the same logic discussed in the previous
sections, i.e.
\begin{eqnarray}
\label{st_correlations}
\left.
\begin{array}{ll}
M_{_{\rm Kerr}}&=M_{_{\rm BH}}=\widetilde{M} \nonumber \\
J_{_{\rm Kerr}}&=J_{_{\rm BH}}
\end{array}
\right\}
\hskip 0.6 cm {\rm internal~orbits},
\\\nonumber\\
\left.
\begin{array}{ll}
M_{_{\rm Kerr}}&=M_{_{\rm tot}}=\widetilde{M} \nonumber \\
J_{_{\rm Kerr}}&=J_{_{\rm tot}}
\end{array}
\right\}
\hskip 0.6 cm {\rm external~orbits}.
\\
\end{eqnarray}
Note that for internal orbits we did try to compare our non-pure Kerr spacetimes with pure Kerr spacetimes having
$M_{_{\rm Kerr}}=M_{_{\rm tot}}=\widetilde{M}$ and $J_{_{\rm Kerr}}=J_{_{\rm tot}}$ (using these values also 
to compute the dephasing time, cf. Sect.~\ref{sec:overlap}), but this turned
out not to be a good choice\footnote{i.e., for many bound stable orbits in
the non-pure Kerr spacetimes that we considered, it was impossible even to
find bound stable orbits with the same
latus rectum and eccentricity in the Kerr spacetime.}.

Once a non-pure and a pure Kerr spacetime have been built and the orbits
have been isolated according to the
relations~\eqref{st_correlations}, further care needs to be paid in
selecting corresponding geodesics.
 As mentioned in
Sect.~\ref{sec:semi-relativistic}, equatorial geodesics can be labelled by two parameters,
which can
be chosen to be, for instance, the latus rectum and the eccentricity $p\mbox{\tiny$_{_{\rm (Q)BL}}$}$ and $e\mbox{\tiny$_{_{\rm (Q)BL}}$}$, calculated 
in QBL coordinates for the non-pure Kerr spacetime and in BL
coordinates for the Kerr spacetime.

However, as already pointed out in
Ref.~\cite{quasi_kerr}, waveforms produced by geodesics having the same 
$p\mbox{\tiny$_{_{\rm (Q)BL}}$}$ and $e\mbox{\tiny$_{_{\rm (Q)BL}}$}$
are significantly different because they do not contain comparable orbital
frequencies, and give overlaps ${\cal O} \lesssim 0.4$. A similar
conclusion can be drawn in the case in which the free parameters are
chosen to be the periastron radius and the (tangential) velocity
measured at the periastron by a zero angular momentum observer (ZAMO):
this choice gives overlaps ${\cal O}\simeq 0.1-0.2$. In view of this,
any sensible comparison can be made only with geodesics in the two
spacetimes that have the same orbital frequencies (this result was
already pointed out in Ref.~\cite{quasi_kerr}).

We recall that an equatorial geodesic in a generic stationary,
axisymmetric spacetime has an $r$-motion that is periodic in the
coordinate time $t$. To see this, it is sufficient to combine
Eqs.~\eqref{eq:geo_t},~\eqref{eq:geo_phi} and the normalization condition
$u_\mu u^\mu=-1$ for an equatorial motion $\theta=\pi/2$ so that
\eq
\label{tmp1}
(dr/dt)^2=V(r,\tilde{\epsilon},\tilde{\ell})\;,
\eeq 
with $V(r,\tilde{\epsilon},\tilde{\ell})$ being a function of $r$ and 
of the two constants of motion $\tilde{\epsilon}$ and $\tilde{\ell}$. Clearly,
Eq.~\eqref{tmp1} has a periodic solution with a frequency that we will
denote $\omega_r$. A similar analysis can be carried out for the
motion in the $\phi$ direction, which, combining Eqs.~\eqref{eq:geo_t}
and~\eqref{eq:geo_phi} with $\theta=\pi/2$, satisfies an equation of
the type
\eq
\label{tmp2}
d\phi/dt=G(r,\tilde{\epsilon},\tilde{\ell})\;,
\eeq 
where $G(r)$ is again a function of $r$, $\tilde{\epsilon}$ and $\tilde{\ell}$. Integrating Eq.~\eqref{tmp2} with
$\phi_0=t_0=0$ leads to
\eq 
\label{tmp3}
\phi(t)=
\langle G\rangle t + \int_0^t (G(r(t),\tilde{\epsilon},\tilde{\ell})-\langle G\rangle)dt\;,
\eeq
where $\langle G\rangle$ is the time average of $G(r(t),\tilde{\epsilon},\tilde{\ell})$ over an
$r$-period. The second term on the right-hand-side of Eq.~\eqref{tmp3} is
clearly periodic (with zero average) in $t$ so that the
$\phi$-motion has a linearly growing term and an oscillating one. The overall frequency content of the $\phi$ motion is 
therefore determined by $\omega_{\phi}=\langle G\rangle$.

The orbital frequencies $\omega_r$ and $\omega_\phi$ can therefore be
used to characterize equatorial geodesics (and hence the waveforms)
that are expected to be as similar as possible (i.e. have the largest
possible overlap) in the two spacetimes. In practice, given a geodesic
(and therefore a waveform) characterized by $\omega_r$ and
$\omega_\phi$ in the non-pure Kerr spacetime, we can compare it to the
waveform produced in the Kerr spacetime by the orbit which has the
same $r$- and $\phi$-frequencies. Since $\omega_r$ and $\omega_\phi$
for equatorial orbits in a Kerr spacetime are functions of $M_{_{\rm
    Kerr}}$, $J_{_{\rm Kerr}}$, $p\mbox{\tiny$_{_{\rm BL}}$}$ and
$e\mbox{\tiny$_{_{\rm BL}}$}$ (explicit expressions for these functions, which we will denote $\omega^{^{\rm Kerr}}_r$ and
$\omega^{^{\rm Kerr}}_\phi$, are given in Ref.~\cite{schmidt}), matching the
geodesics amounts to solving the following equations in the unknowns
$\delta p$ and $\delta e$
\begin{eqnarray}
&&\omega^{^{\rm BH+Torus}}_r(p\mbox{\tiny$_{_{\rm
        QBL}}$},e\mbox{\tiny$_{_{\rm
        QBL}}$})=
  \nonumber\\ &&\omega^{^{\rm Kerr}}_r(p\mbox{\tiny$_{_{\rm BL}}$}\!=
  \!p\mbox{\tiny$_{_{\rm QBL}}$}+\delta p,e\mbox{\tiny$_{_{\rm
        BL}}$}\!=\!e\mbox{\tiny$_{_{\rm QBL}}$}+\delta e,M_{_{\rm
      Kerr}},J_{_{\rm Kerr}})\;, 
\label{eq:same_per_1} \nonumber\\\\
&&\omega^{^{\rm BH+Torus}}_\phi(p\mbox{\tiny$_{_{\rm
      QBL}}$},e\mbox{\tiny$_{_{\rm
      QBL}}$})=
\nonumber\\ &&\omega^{^{\rm Kerr}}_\phi(p\mbox{\tiny$_{_{\rm BL}}$}\!=
\!p\mbox{\tiny$_{_{\rm QBL}}$}+\delta p,e\mbox{\tiny$_{_{\rm
      BL}}$}\!=\!e\mbox{\tiny$_{_{\rm QBL}}$}+\delta e,M_{_{\rm
    Kerr}},J_{_{\rm Kerr}}) \,,  \nonumber\\
 \label{eq:same_per_2}
\end{eqnarray}
where $\omega^{^{\rm BH+Torus}}_r(p\mbox{\tiny$_{_{\rm QBL}}$},e\mbox{\tiny$_{_{\rm QBL}}$})$ and 
$\omega^{^{\rm BH+Torus}}_\phi(p\mbox{\tiny$_{_{\rm QBL}}$},e\mbox{\tiny$_{_{\rm QBL}}$})$ 
are the $r$- and $\phi$-frequencies of 
the equatorial orbit with latus rectum $p\mbox{\tiny$_{_{\rm QBL}}$}$ and eccentricity $e\mbox{\tiny$_{_{\rm QBL}}$}$ in the non-pure Kerr spacetime under consideration and
where $M_{_{\rm Kerr}},J_{_{\rm Kerr}}$ follow the selection rule in
Eq.~\eqref{st_correlations} to distinguish internal and external orbits.
Indeed, this is the approach which was followed in Ref.~\cite{quasi_kerr} and
which highlighted the possibility of a confusion problem in non-pure
Kerr spacetimes. 

An important difference with respect to the work presented in
Ref.~\cite{quasi_kerr} is that we also considered a different way in which 
it is possible to identify geodesics that have the same orbital frequencies in a Kerr and in a
non-pure Kerr spacetime. We can in fact
consider the latus rectum and eccentricity fixed in (Q)BL and search for the values of the additional mass
$\delta M$ and angular momentum $\delta J$ of the Kerr spacetime which
would yield the same $r$- and $\phi$-frequencies, i.e. 
\begin{eqnarray}
&&\omega^{^{\rm BH+Torus}}_r(p\mbox{\tiny$_{_{\rm
        QBL}}$},e\mbox{\tiny$_{_{\rm
        QBL}}$})=
  \nonumber\\ &&\omega^{^{\rm Kerr}}_r(p\mbox{\tiny$_{_{\rm BL}}$}\!=
  \!p\mbox{\tiny$_{_{\rm QBL}}$},e\mbox{\tiny$_{_{\rm
        BL}}$}\!=\!e\mbox{\tiny$_{_{\rm QBL}}$},M_{_{\rm
      Kerr}}+\delta M,J_{_{\rm Kerr}}+\delta J)\;, 
\label{eq:same_per_3} \nonumber\\\\
&&\omega^{^{\rm BH+Torus}}_\phi(p\mbox{\tiny$_{_{\rm
      QBL}}$},e\mbox{\tiny$_{_{\rm
      QBL}}$})=
\nonumber\\ &&\omega^{^{\rm Kerr}}_\phi(p\mbox{\tiny$_{_{\rm BL}}$}\!=
\!p\mbox{\tiny$_{_{\rm QBL}}$},e\mbox{\tiny$_{_{\rm
      BL}}$}\!=\!e\mbox{\tiny$_{_{\rm QBL}}$},M_{_{\rm
    Kerr}}+\delta M,J_{_{\rm Kerr}}+\delta J) \,.  \nonumber\\
 \label{eq:same_per_4}
\end{eqnarray}
Of course, a similar but distinct set of equations can also be built
by considering orbits having the same latus rectum and eccentricity in
QI coordinates\footnote{The transformation from BL to QI coordinates
in a Kerr spacetime is given for instance in Ref.~\cite{cook}, Eq.~(80):
the transformation turns out to be the inverse of Eq.~\eqref{eq:QI2QBL}, with
$r_{_{+\,\rm QI}}=M(1-a^2)^{1/2}/2$ ($M$ and $a$ being
the mass and the spin parameter of the Kerr spacetime under consideration).}
\begin{eqnarray}
&&\omega^{^{\rm BH+Torus}}_r(p\mbox{\tiny$_{_{\rm
        QI}}$},e\mbox{\tiny$_{_{\rm
        QI}}$})=
  \nonumber\\ &&\omega^{^{\rm Kerr}}_r(p\mbox{\tiny$_{_{\rm QI}}$},e\mbox{\tiny$_{_{\rm QI}}$},M_{_{\rm
      Kerr}}+\delta M,J_{_{\rm Kerr}}+\delta J)\;, 
\label{eq:same_per_5} 
\end{eqnarray}
\begin{eqnarray}
&&\omega^{^{\rm BH+Torus}}_\phi(p\mbox{\tiny$_{_{\rm
      QI}}$},e\mbox{\tiny$_{_{\rm
      QI}}$})=
\nonumber\\ &&\omega^{^{\rm Kerr}}_\phi(p\mbox{\tiny$_{_{\rm QI}}$},e\mbox{\tiny$_{_{\rm
      QI}}$},M_{_{\rm Kerr}}+\delta M,J_{_{\rm Kerr}}+\delta J) \,.
 \label{eq:same_per_6}
\end{eqnarray}

\begin{table*}[tbh]
\begin{tabular}{|c|c|c|c|c|c|c|c|c|c|c|c|c|c|}
\hline 
 \T \B spacetime &$M_{_{\rm BH}}$&$M_{_{\rm Torus}}$&${M_{_{\rm BH}}}/{M_{_{\rm Torus}}}$&$J_{_{\rm BH}}$&$J_{_{\rm Torus}}$& ${J_{_{\rm BH}}}/{J_{_{\rm Torus}}}$&${J_{_{\rm BH}}}/{{M^2_{_{\rm BH}}}}$ &${J_{_{\rm tot}}}/{{M^2_{_{\rm tot}}}}$&$\rho$&$r_{_{+\,,\rm QI}}$&$r_{_{\rm in\,, QI}}$&$r_{_{\rm out\,, QI}}$&$\epsilon$\\
\hline
 \T \B $A$& 0.413& 0.121& 3.4& $+9.02{\rm e-}2$& $1.17{\rm
   e-}1$&$+7.69{\rm e-}1$&$+5.28{\rm e-}1$& 0.728& 2.637 &0.179& 0.6064& 0.6305& 0.11\\
\hline
\T \B $B$& 0.100& 0.007& 14.3 & $-1.74{\rm e-}5$ & $2.58{\rm e-}3$ & $-6.74{\rm e-}3$& $-1.74{\rm e-}3$& 0.224 &0.198&  0.050  & 0.9156 & 1.0000 & 2.63\\
\hline
\end{tabular}
\caption{Parameters of the spacetimes analyzed in section
  \ref{sec:comparison}, in units in which $10^7 M_\odot=G=c=1$. $r_{_{\rm in\,, QI}}$ and $r_{_{\rm
      out\,, QI}}$ are the inner and outer edges of the torus in QI
  coordinates, $\rho$ is the baryonic mass density of the torus and the parameter $\epsilon$ provides a lowest-order
  measure of the deviation of the spacetime away from a Kerr solution
  [cf. Eq.~\eqref{eq:eps}]. Note that $\epsilon$ is more sensitive to
  the ratio between the angular momenta than to that between the
  masses.}\label{table:spacetimes_data}
\end{table*}

\begin{figure}
\includegraphics[width=8.5cm, angle=0]{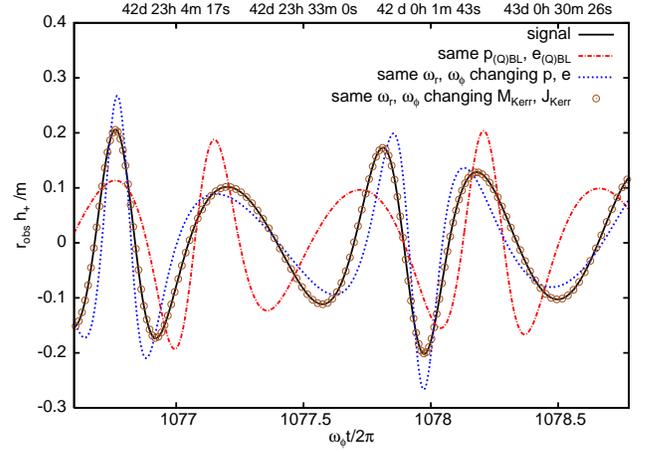}
\caption{Kludge waveforms around the dephasing time for a small body with mass $m=1M_\odot$ moving in the spacetime $B$ of Table~\ref{table:spacetimes_data}. The black solid line
  shows the waveform produced by a geodesic with given latus rectum
  and eccentricity in spacetime $B$, while the red dot-dashed
  one refers to a geodesic with the same latus rectum and eccentricity
  (in (Q)BL coordinates) in a Kerr spacetime with $M_{_{\rm
      Kerr}}=M_{_{\rm tot}}$ and $J_{_{\rm Kerr}}=J_{_{\rm tot}}$. The
  blue dotted line and the brown circles are instead the waveforms
  produced by an orbit with the same $r$- and $\phi$-frequencies as
  obtained by adjusting $(\delta p, \delta e)$ or $(\delta M, \delta
  J)$, respectively.
\label{fig:tone1}}
\end{figure}

To illustrate how different correlations of orbits in the two
spacetimes can lead to significantly different waveforms, we show in
Fig.~\ref{fig:tone1} some kludge waveforms 
for a small body with mass $m=1M_\odot$ moving in the spacetime $B$ whose parameters are listed in Table~\ref{table:spacetimes_data}.
The geodesics have been calculated up to the dephasing time
(i.e. $\tau_{\rm d} \simeq 42$ d) and the figure shows a magnification
of the waveforms around this time. In particular, the black solid line
shows the waveform produced by a geodesic with $p_{_{\rm
    QBL}}/M_{_{\rm tot}}=21.237$ and $e_{_{\rm QBL}}=0.212$ in 
spacetime $B$, while the red dot-dashed one refers to a geodesic
with the same latus rectum and eccentricity (in (Q)BL coordinates) in a
Kerr spacetime with $M_{_{\rm Kerr}}=M_{_{\rm tot}}$ and $J_{_{\rm
    Kerr}}=J_{_{\rm tot}}$. The blue dotted line and the brown circles
are instead the waveforms produced by an orbit with the same $r$- and
$\phi$-frequencies as obtained by adjusting $(\delta p, \delta e)$ or
$(\delta M, \delta J)$, respectively. Clearly, fixing the same orbital
parameters $p_{_{\rm (Q)BL}}$ and $e_{_{\rm (Q)BL}}$ in the two spacetimes would be misleading and
will inevitably produce very small overlaps. On the other hand,
ensuring that the orbital frequencies are the same by adjusting $\delta M$ and $\delta
J$ provides waveforms that are much more similar and
even harder to distinguish over this timescale than if $\delta
p$ and $\delta e$ are adjusted.

In the following sections we will discuss in detail the confusion
problem when considering the two different ways in which the geodesics
in the two spacetimes can be matched. Before doing that, however, we will
now briefly recall the main properties of the numerically-generated
spacetimes that we have considered here, and whose parameters are
listed in Table~\ref{table:spacetimes_data}. We note that because the
investigation of each spacetime is a rather lengthy and computationally
expensive operation, we have restricted our attention to two
spacetimes only, but with rather different properties. More
specifically, we have considered a first spacetime (denoted as $A$)
having a torus with mass comparable with that of the black hole and slightly
larger angular momentum (i.e. $M_{_{\rm BH}} \gtrsim M_{_{\rm Torus}},
\vert J_{_{\rm BH}}\vert \lesssim\vert J_{_{\rm Torus}}\vert$) and a second spacetime
(denoted as $B$) having a torus with mass much smaller than that of
the black hole but much larger angular momentum (i.e. $M_{_{\rm BH}}
\gg M_{_{\rm Torus}}, \vert J_{_{\rm BH}}\vert \ll \vert J_{_{\rm Torus}}\vert$). 

We also note that spacetime $A$ has a rather small quadrupole parameter
$\epsilon \simeq 0.1$ [cf. Eq.~\eqref{eq:eps} for the definition] and
could therefore be used to validate the perturbative results of
Ref.~\cite{quasi_kerr} which, we recall, were formulated to the lowest
order in $\epsilon$.  Interestingly, we will see that taking into account
the higher-order multipoles can lead to important qualitative differences
and weaken or even cancel, for orbits very close to the torus, the
confusion problem found in Ref.~\cite{quasi_kerr}.  Spacetime $B$, on the
other hand, has a considerably larger value for $\epsilon$ and cannot,
therefore, be described satisfactorily by the
metric~\eqref{eq:quasi_kerr}. The spacetimes were computed to
sufficiently high accuracy so as to ensure that the numerical errors do
not affect the results. More specifically, for spacetime $A$ we used $40
\times 40$ Chebyshev polynomials in the vacuum domain extending out to
infinity (domain 1 of \cite{BH_plus_ring}) and $28 \times 28$ polynomials
in the other 4 domains. For spacetime $B$ we used $31 \times 27$
polynomials in each of the 5 domains. Typical physical quantities, such
as mass and angular momentum, were thus accurate to about $10^{-6}$ in
spacetime $A$ and $10^{-7}$ in spacetime $B$. Besides these being errors
that are orders of magnitude smaller than the ones
${O}(a\,\epsilon,\,\epsilon^2) + {O} (\delta M_{\ell\geq4}, \delta
S_{\ell\geq3})$ typically affecting the approximate
metric~\eqref{eq:quasi_kerr}, the accuracy of our numerically generated
spacetimes is sufficient for our purposes, since the dephasing it
introduces is comparable with the dephasing due to radiation reaction, as
the latter scales with the mass ratio $m/M_{\rm
  BH}\approx10^{-6}-10^{-7}$. As a result, introducing a cut-off at the
dephasing time not only makes the effects of radiation reaction
negligible, but it also ensures that the numerical errors in the
calculation of the spacetimes do not affect the results. As a further
check, we have varied the number of Chebyshev polynomials and verified
that the numerical errors inherent to the spacetimes have a negligible
impact on our final results.

For all of the waveforms computed in this paper, we have considered an
observer located at $\phi_{obs}=0, \theta_{obs}=\pi/4$ and decomposed the
incoming gravitational-wave signal into the usual ``plus'' and ``cross''
polarizations (see, for instance, Refs.~\cite{babak_kludge,MTW} for
details). Furthermore, labelling the gravitational waves computed in the
two spacetimes with 1 and 2, we calculate the overlap between both
polarizations, ${\cal O}(h_1^+,h_2^+)$ and ${\cal
  O}(h_1^\times,h_2^\times)$, and in the discussion of our results we
refer to the smallest of the two overlaps, i.e. ${\cal
  O}(h_1,h_2)\equiv\min[{\cal O}(h_1^+,h_2^+),\,{\cal
  O}(h_1^\times,h_2^\times)]$. In practice, however, the difference
between ${\cal O}(h_1^+,h_2^+)$ and ${\cal O}(h_1^\times,h_2^\times)$ for
the overlaps plotted in the figures of the next sections is typically
smaller than 0.005 and in no case larger than 0.025.

Finally, we note that the results presented in the next sections refer to
a small body with $m=1M_\odot$ and to a sensitivity function for LISA
computed using the online generator~\cite{sensitivity} with its default
parameters and, in particular, no white dwarf noise. As pointed out in
Ref.~\cite{quasi_kerr}, including white-dwarf noise would only lead to a
slight increase in the dephasing time.

%===============================================
\subsection{The confusion problem when varying $e$ and $p$}
\label{sec:res1}
%===============================================

After excluding a comparison between geodesics (and hence waveforms) that
have the same latus rectum and eccentricity in the pure and non-pure Kerr
spacetimes because of the very small overlap they produce, we have
compared waveforms having the same $r$- and $\phi$-frequencies as
obtained by changing the latus rectum and eccentricity while keeping
$M_{_{\rm Kerr}}$ and $J_{_{\rm Kerr}}$ fixed [cf.
Eqs.~\eqref{eq:same_per_1} and~\eqref{eq:same_per_2}].  More
specifically, we already mention that the values of $\delta p/p_{_{\rm
    QBL}}$ obtained in the regions of the space of parameters $(p_{_{\rm
    QBL}},\,e_{_{\rm QBL}})$ where the overlap between these waveforms is
high (${\cal O}>0.95$) are $\vert\delta p/p_{_{\rm QBL}}\vert \lesssim
0.05$ in spacetime $A$ and $\vert\delta p/p_{_{\rm QBL}}\vert \lesssim
0.16$ in spacetime $B$. Similarly, the values of $\delta e$ obtained in
the regions of the space of parameters where ${\cal O}>0.95$ are
$\vert\delta e\vert\lesssim 0.06$ in spacetime $A$ and $\vert\delta
e\vert\lesssim 0.07$ in spacetime $B$.

\begin{figure*}[tbh]
\includegraphics[width=8.2cm, angle=-90]{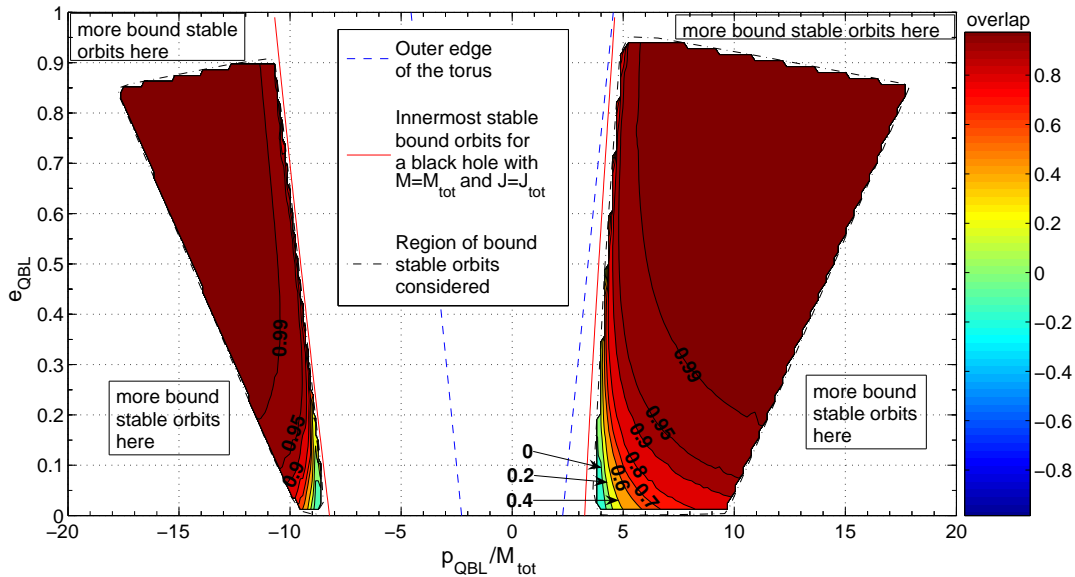}
\caption{Overlap between waveforms produced in spacetime $A$ by {\sl
    external} orbits and waveforms produced in a Kerr spacetime with
  mass $M_{_{\rm Kerr}}=M_{_{\rm tot}}$ and spin $J_{_{\rm Kerr}}=J_{_{\rm tot}}$. 
 The orbits all have
  the same $r$- and $\phi$-frequencies as obtained by suitably changing
  the latus rectum and the eccentricity, with positive values of
  $p_{_{\rm QBL}}$ referring to prograde orbits, and negative ones to
  retrograde orbits. The different lines mark the margins of the
  different relevant regions of the $(p_{_{\rm QBL}},\,e_{_{\rm QBL}})$ plane, with the blue dashed line
  representing the outer ``edge of the torus'', the red solid line
  representing the innermost stable bound orbits for a Kerr spacetime
  with mass $M_{_{\rm Kerr}}=M_{_{\rm tot}}$ and spin $J_{_{\rm Kerr}}=J_{_{\rm tot}}$ and the black
  dot-dashed line limiting the regions of the $(p_{_{\rm QBL}},\,e_{_{\rm QBL}})$ plane where
  bound stable orbits have been studied. A high overlap in large regions of the   space of parameters indicates that a confusion problem is indeed
  possible in this spacetime for observational timescales below or comparable to the
  dephasing time, although this confusion disappears for orbits with small eccentricities and close to the innermost bound stable orbits.}
\label{fig:overlap_conf_pe_BH3}
\vskip 0.250cm
\end{figure*}
\begin{figure*}[tbh]
\includegraphics[width=8.2cm, angle=-90]{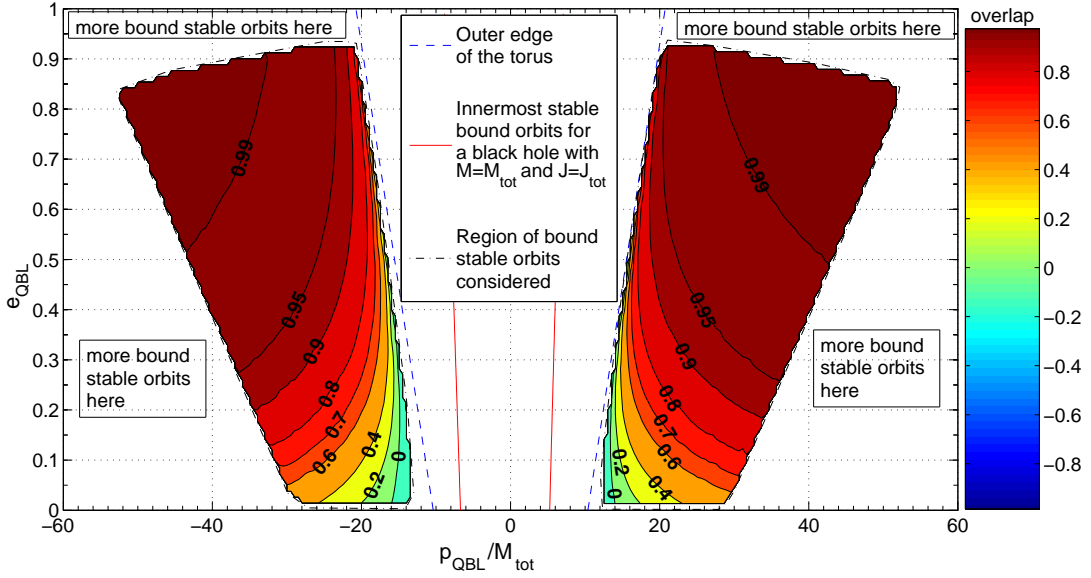}
\caption{The same as in Fig.~\ref{fig:overlap_conf_pe_BH3} but for
  spacetime $B$. Note that in this case the confusion problem is less
  severe and indeed not present for orbits near the outer
edge of the torus (i.e., with $p_{_{\rm QBL}}/M_{_{\rm tot}}\lesssim 30$) and
with eccentricities $e_{_{\rm QBL}} \lesssim 0.2$.
\label{fig:overlap_conf_pe_BH14ext}}
\vskip 0.250cm
\end{figure*}
\begin{figure*}[tbh]
\includegraphics[width=8.2cm, angle=-90]{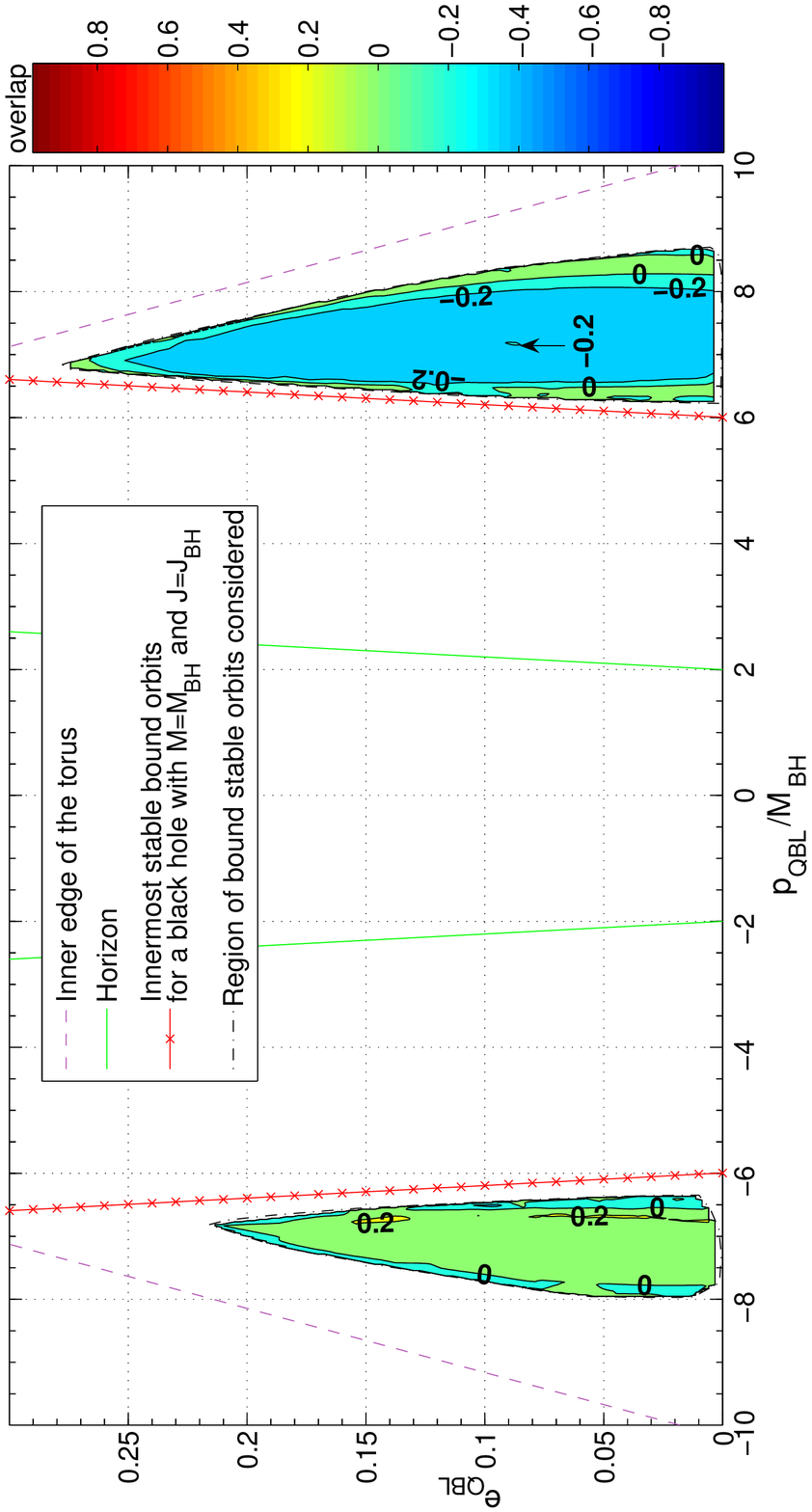}
\caption{The same as in Fig.~\ref{fig:overlap_conf_pe_BH14ext} but for
  {\sl internal} orbits, with the green solid line marking those orbits whose periastron
lies on the event horizon, the purple dashed line representing the inner ``edge of the torus'' and finally the red crossed-solid line marking the innermost bound stable orbits in a
Kerr spacetime with mass and spin $M_{_{\rm Kerr}}=M_{_{\rm BH}}$ and $J_{_{\rm Kerr}}=J_{_{\rm BH}}$. Again, the black dot-dashed line limits the regions of the $(p_{_{\rm QBL}},\,e_{_{\rm QBL}})$ plane where bound stable orbits have been studied, but in contrast to the case of external orbits, these regions correspond to practically all the bound stable orbits not crossing the torus.
Note that in this case
  the confusion problem is absent, with ${\cal O} \lesssim 0.2$. \label{fig:overlap_conf_pe_BH14int}}
\vskip 0.250cm
\end{figure*}

We have computed the overlap between
$h_{_{\rm BH+Torus}}(p_{_{\rm QBL}}, e_{_{\rm QBL}})$ and $h_{_{\rm
    Kerr}}(p_{_{\rm BL}}=p_{_{\rm QBL}} +\delta p, e_{_{\rm
    BL}}=e_{_{\rm QBL}}+\delta e, M_{_{\rm Kerr}}, J_{_{\rm Kerr}})$,
and summarize the results of this comparison for a large set of orbits
in Figs.~\ref{fig:overlap_conf_pe_BH3},
\ref{fig:overlap_conf_pe_BH14ext} and
\ref{fig:overlap_conf_pe_BH14int}. In particular,
Fig.~\ref{fig:overlap_conf_pe_BH3} shows the color-coded overlap
between waveforms produced in spacetime $A$ by external orbits in a
$(p_{_{\rm QBL}},\,e_{_{\rm QBL}})$ plane, with positive values of
$p_{_{\rm QBL}}$ referring to prograde orbits, and negative ones to
retrograde orbits. Note that {\textit no} internal orbits were found in
spacetime $A$ and this is due to the fact that in this case the torus
is too close to the black hole for bound stable orbits to exist in
region I of Fig.~\ref{spacetime_regions} without plunging into the black hole.
 The different lines in Fig.~\ref{fig:overlap_conf_pe_BH3} mark
the margins of the different regions of interest in the $(p_{_{\rm QBL}},\,e_{_{\rm QBL}})$
plane, with the blue dashed line representing the outer ``edge of the
torus'', that is the set of points such that $p_{_{\rm QBL}}/(1+e_{_{\rm QBL}})=r_{_{\rm out,
    QBL}}$. Similarly, the red solid line represents the innermost
stable bound orbits (this line is also referred to as the
``separatrix'' in Ref.~\cite{separatrix}) for a Kerr spacetime with
mass $M_{_{\rm Kerr}}=M_{_{\rm tot}}$ and spin $J_{_{\rm Kerr}}=J_{_{\rm tot}}$. Finally, the black
dot-dashed line limits the regions of the
$(p_{_{\rm QBL}},\,e_{_{\rm QBL}})$ plane where bound stable orbits have been studied.

We underline that these are not the only regions in which bound stable
orbits exist, but they rather represent the regions we have investigated
because of their being more directly related to LISA observations. In
practice, we exploit the fact that there is a one-to-one correspondence
between the latus rectum $p_{_{\rm QBL}}$ and the eccentricity $e_{_{\rm
QBL}}$ of bound stable orbits and their QI radius and tangential velocity
(measured by a ZAMO) at periastron, $r_p$ and $v_\phi$.  We therefore
choose the initial radial QI position $r_0$ of the small body randomly in
a limited range and vary its initial tangential velocity $v_\phi$ with
small steps in the range of the velocities leading to energies per unit
mass $\tilde\epsilon<1$.\footnote{We note that in both spacetime $A$ and
$B$ all the equatorial bound stable orbits not crossing the torus have
$\tilde{\epsilon}<1$ [this can be verified by computing the values of
$\tilde{\epsilon}$ for which the potential
$V(r,\tilde{\epsilon},\tilde{\ell})$ in Eq.~\eqref{tmp1} is positive].
However, bound stable orbits which cross the torus and have
$\tilde{\epsilon}>1$ are present in both spacetimes.} After integrating
the geodesic equations (Eqs.~\eqref{eq:geo_t}-\eqref{eq:geo_r} with
$\theta=\pi/2)$ over and beyond the dephasing time, if the orbit does
not intersect the torus and if $r_0$ actually corresponds to the
periastron (and \textit{not} to the apoastron) we extract the latus
rectum and eccentricity so as to populate the $(p_{_{\rm QBL}},\,e_{_{\rm
QBL}})$ plane and compute the overlaps with pure-Kerr waveforms (the
orbits in the Kerr spacetime are chosen to start at their periastron as
well).  Overall, a large number of bound stable orbits (i.e. $\gtrsim
2250$) has been integrated for each of the figures shown in this
paper. Notice that the requirement that $r_0$ correspond to the
periastron is important because, as far as the overlaps are concerned,
orbits having the same latus rectum and eccentricity but different
initial positions are not equivalent. We recall in fact that the overlaps
are computed by putting a cutoff at the dephasing time and if the initial
positions are different, the portions of the orbits contributing to the
overlap are different.

Overall, because the waveforms agree very well with an overlap
${\cal O}> 0.95$ for most of the orbits we have considered, the results in
Fig.~\ref{fig:overlap_conf_pe_BH3} clearly show that a confusion
problem similar to the one presented in Ref.~\cite{quasi_kerr} is
indeed possible in this spacetime for observational timescales below or comparable to
the dephasing time. As indicated by the color-coding, the overlap has
a drastic reduction only in a limited region of the space of
parameters and in particular for orbits with small eccentricity and
close to the innermost bound stable orbits. This is not surprising as in
these regions the local modifications of the spacetime due to the
presence of the torus are the largest and have a more marked impact on
the waveforms. Interestingly, prograde orbits produce
overlaps that are smaller than those produced by retrograde orbits with comparable 
values of $p_{_{\rm QBL}}$ and $e_{_{\rm QBL}}$, and appear therefore to
be better tracers of this spacetime.

It is important to underline that the presence of an albeit small
region of the space of parameters in which the overlap is small, and
hence the dangers of a confusion problem decreased, represents an
important difference compared to the results presented in
Ref.~\cite{quasi_kerr}. We recall that spacetime $A$ has a rather
small quadrupole parameter $\epsilon$
(cf. Table~\ref{table:spacetimes_data}), comparable with those used
in Ref.~\cite{quasi_kerr}. Yet, the small overlaps near the innermost
 bound stable orbits indicate that taking into account the higher-order multipoles neglected in
the metric~\eqref{eq:quasi_kerr} can lead to significant differences
even far away from the black hole if a matter source is present.

Figure~\ref{fig:overlap_conf_pe_BH14ext} summarizes a set of results
similar to those presented in Fig.~\ref{fig:overlap_conf_pe_BH3} but
for spacetime $B$. More specifically, it reports the color-coded
overlap between waveforms produced in spacetime $B$ by {\sl external}
orbits and waveforms produced in a Kerr spacetime with mass $M_{_{\rm
    Kerr}}=M_{_{\rm tot}}$ and spin $J_{_{\rm Kerr}}=J_{_{\rm tot}}$. Here again, all of the orbits
have the same orbital frequencies as obtained by adjusting $\delta p$ and
$\delta e$. It should be noted that in this case the confusion problem is
less severe and indeed essentially absent for orbits near the outer
edge of the torus (i.e., with $p_{_{\rm QBL}}/M_{_{\rm tot}}\lesssim 30$) and
with eccentricities $e_{_{\rm QBL}} \lesssim 0.2$. Finally, we
report in Fig.~\ref{fig:overlap_conf_pe_BH14int} again results for
spacetime $B$ but this time for {\sl internal} orbits. We recall, in
fact, that in this case the torus is farther away from the black hole
and thus bound stable orbits can be found in region I of
Fig.~\ref{spacetime_regions}. 
As in the previous figures, the black dot-dashed line limits the regions of the plane $(p_{_{\rm QBL}},\,e_{_{\rm QBL}})$
where bound stable orbits have been studied, 
but in contrast to the case of external orbits these regions correspond to practically all the bound stable orbits not crossing the torus.
On the other hand, the green solid line marks those orbits whose periastron
lies on the event horizon, the purple dashed one those orbits whose apoastron
lies on the inner edge of the torus and finally the red crossed-solid line indicates the innermost bound stable orbits in a
Kerr spacetime with mass and spin $M_{_{\rm Kerr}}=M_{_{\rm BH}}$ and $J_{_{\rm Kerr}}=J_{_{\rm BH}}$.
Clearly, no confusion problem is present
for these orbits, because the overlap is always very small and never larger
than $\simeq 0.2$.

In summary, the overlap computed in the two spacetimes $A$ and $B$ containing a
black-hole and a torus by varying the latus rectum and the
eccentricity reveals that there are regions in which the non-pure Kerr spacetimes
can be ``confused'' with Kerr spacetimes that are equivalent to them at the
sensitivity of LISA. Clearly, this risk is concrete only for
timescales over which radiation-reaction effects are negligible and it
is not present for external orbits very close to the torus 
or for the orbits between the torus and the black hole, if they exist.

 \begin{figure*}
 \includegraphics[width=8cm, angle=-90]{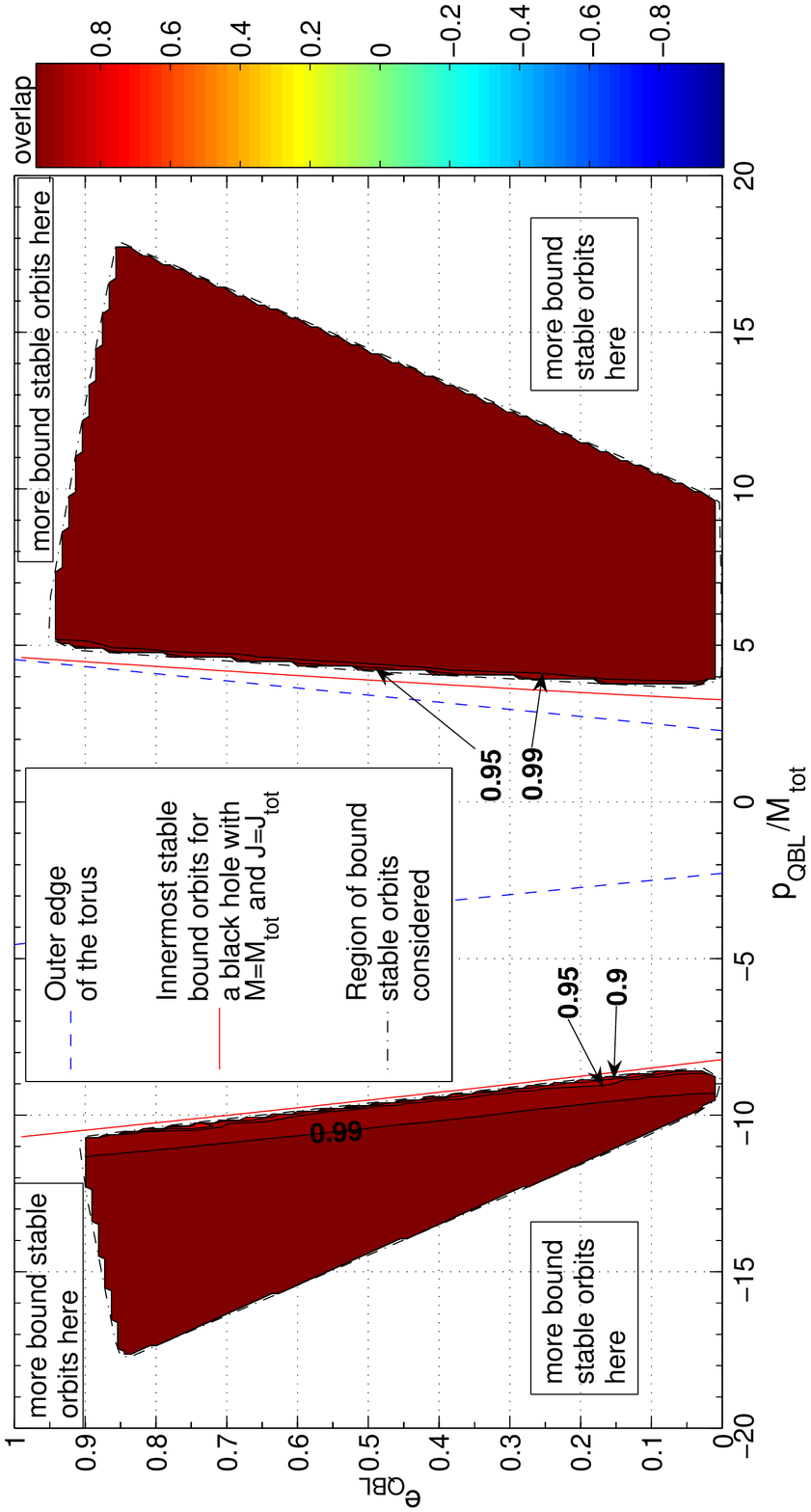}
 \caption{Overlap between waveforms produced in spacetime $A$ by {\sl
    external} orbits and waveforms produced in a Kerr spacetime with
  mass $M_{_{\rm Kerr}}+\delta M=M_{_{\rm tot}}+\delta M$ and spin $J_{_{\rm Kerr}}+\delta J=J_{_{\rm tot}}+\delta J$ by
  orbits with the same latus rectum and eccentricity [in (Q)BL
    coordinates] and the same $r$- and $\phi$-frequencies. Here too,
  the blue dashed line represents the outer ``edge of the torus'', the
  red solid line the innermost stable bound orbits for a Kerr
  spacetime with mass $M_{_{\rm Kerr}}=M_{_{\rm tot}}$ and spin $J_{_{\rm Kerr}}=J_{_{\rm tot}}$ and
  the black dot-dashed line limits the regions of the $(p_{_{\rm QBL}},\,e_{_{\rm QBL}})$
  plane where bound stable orbits have been studied. An overlap ${\cal
    O}> 0.95$ is present in all of the relevant regions of the $(p_{_{\rm QBL}},\,e_{_{\rm QBL}})$ plane.}
  \label{fig:overlap_conf_am_BH3}
\vskip 0.250cm
\end{figure*}
\begin{figure*}
\vskip 0.250cm
\includegraphics[width=7.9cm, angle=-90]{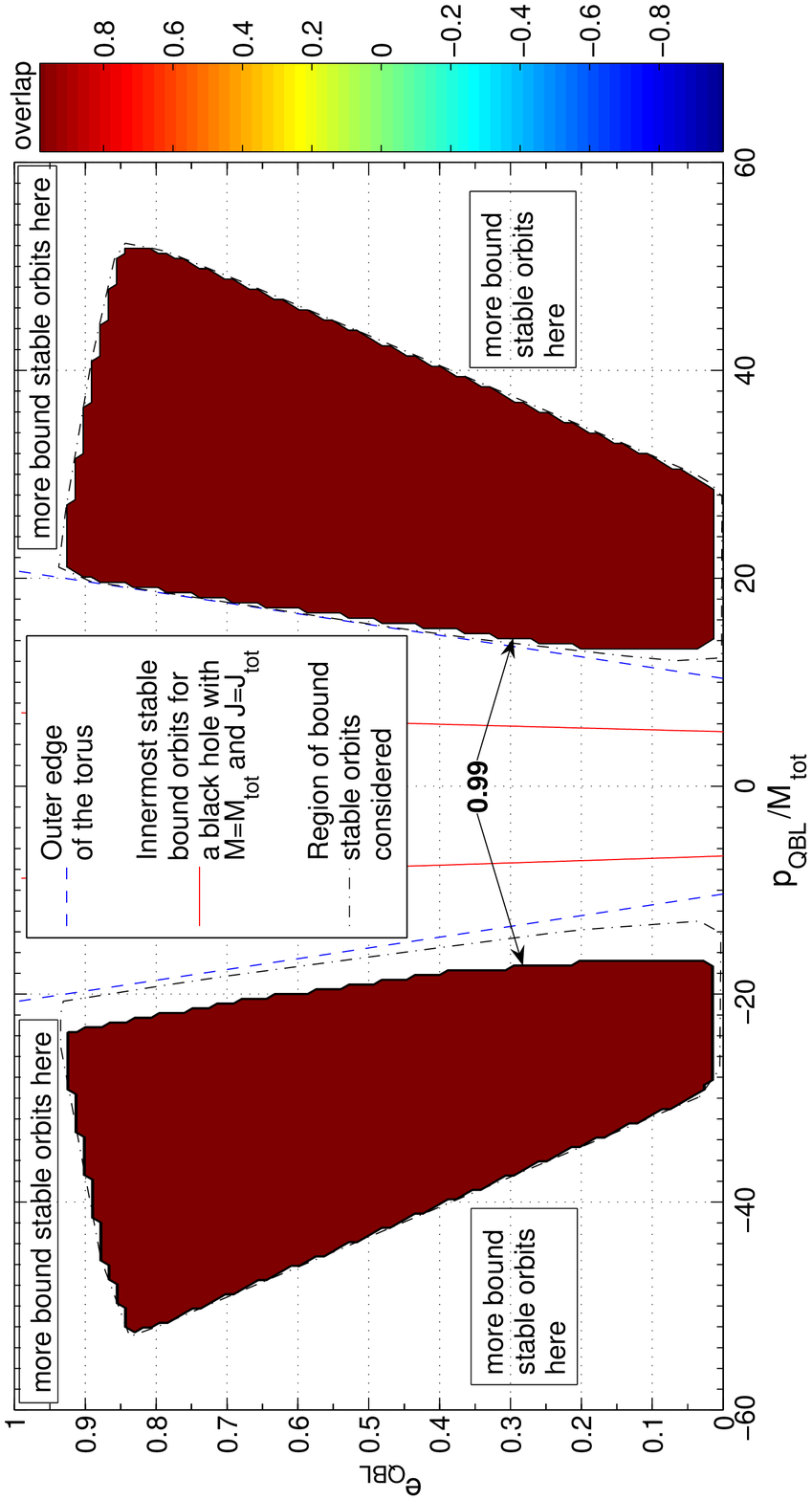}
\caption{The same as in Fig.~\ref{fig:overlap_conf_am_BH3} but for
  spacetime $B$. Note that also in this case the overlap is very high (${\cal O}> 0.99$)
  in almost all of the relevant regions of the $(p_{_{\rm QBL}},\,e_{_{\rm QBL}})$ plane, with the exception of 
  a very small set of orbits very close to the torus, for which Eqs.~\eqref{eq:same_per_3}--\eqref{eq:same_per_4}
  have no solutions (these orbits correspond to the blank regions inside 
  the black dot-dashed line).}
\label{fig:overlap_conf_am_BH14ext}
\end{figure*}

%===============================================
\subsection{The confusion problem when varying $M$ and $J$}
\label{sec:res2}
%===============================================

Next, we consider the overlap obtained by comparing orbits having
the same $r$- and $\phi$-frequencies, which was achieved by changing the mass
and spin of the Kerr black hole while keeping the latus rectum and
eccentricity fixed in either (Q)BL or QI coordinates
[cf. Eqs.~\eqref{eq:same_per_3}--\eqref{eq:same_per_4}
  and~\eqref{eq:same_per_5}--\eqref{eq:same_per_6}]. Doing this
corresponds to considering a hypothetical scenario in which it would be possible to
measure, through independent astronomical observations, the latus
rectum and eccentricity of the small body orbiting around the massive
central black hole. In practice, and using the same compact notation
introduced above, we have compared waveforms of the type $h_{_{\rm
    BH+Torus}}(p_{_{\rm QBL}}, e_{_{\rm QBL}})$ with $h_{_{\rm
    Kerr}}(p_{_{\rm BL}}=p_{_{\rm QBL}}, e_{_{\rm BL}}=e_{_{\rm QBL}},
M_{_{\rm Kerr}} + \delta M, J_{_{\rm Kerr}} + \delta J)$ [i.e., latus
  rectum and eccentricity fixed in (Q)BL coordinates, $\delta M$ and
  $\delta J$ solutions to
  Eqs.~\eqref{eq:same_per_3}--\eqref{eq:same_per_4}] and $h_{_{\rm
    BH+Torus}}(p_{_{\rm QI}}, e_{_{\rm QI}})$ to $h_{_{\rm
    Kerr}}(p_{_{\rm QI}}, e_{_{\rm QI}}, M_{_{\rm Kerr}}+\delta
M,J_{_{\rm Kerr}}+\delta J)$ [i.e., latus rectum and eccentricity fixed
  in QI coordinates, $\delta M$ and $\delta J$ solutions to
  Eqs.~\eqref{eq:same_per_5}--\eqref{eq:same_per_6}]. 

\begin{figure*}
\vskip 0.5cm
\includegraphics[width=8.3cm, angle=-90]{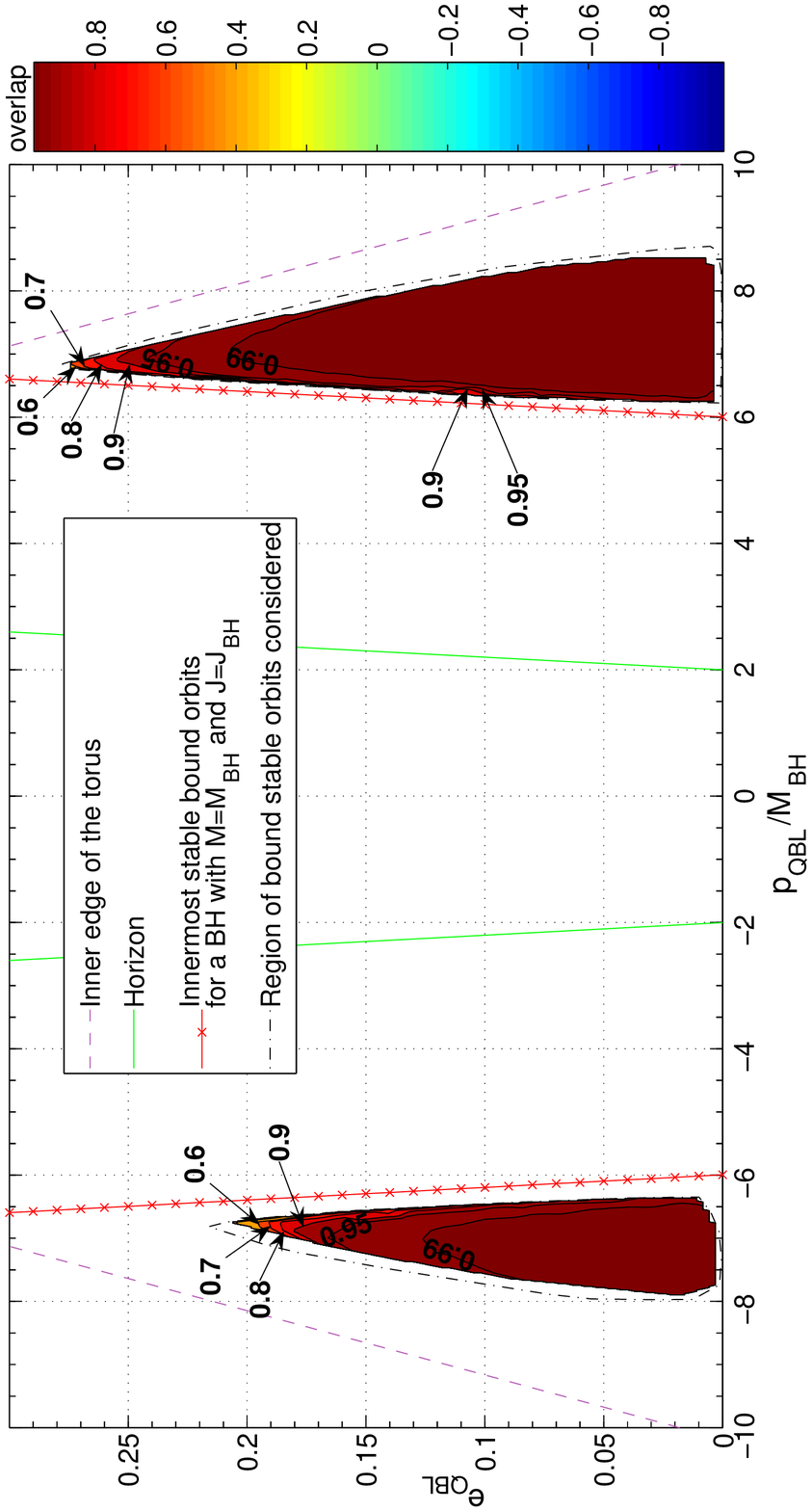}
\caption{The same as in Fig.~\ref{fig:overlap_conf_am_BH14ext} but for
  {\sl internal} orbits, with the green solid line marking those orbits whose periastron
lies on the event horizon, the purple dashed line representing the 
inner ``edge of the torus'' and finally the red crossed-solid line marking the innermost bound stable orbits in a
Kerr spacetime with mass and spin $M_{_{\rm Kerr}}=M_{_{\rm BH}}$ and $J_{_{\rm Kerr}}=J_{_{\rm BH}}$.  Again, the black dot-dashed line limits the regions of the $(p_{_{\rm QBL}},\,e_{_{\rm QBL}})$ plane where bound stable orbits have been studied, but in contrast to the case of external orbits these regions correspond to practically all the bound stable orbits not crossing the torus.
 Note that in this case the confusion
  problem is present in most of the relevant regions of the $(p_{_{\rm QBL}},\,e_{_{\rm QBL}})$
  plane, becoming slightly less severe only for the largest allowed
  eccentricities and for a very small set of orbits, very close to the torus, for which Eqs.~\eqref{eq:same_per_3}--\eqref{eq:same_per_4}
  have no solutions (these orbits correspond to the blank regions inside 
the black dot-dashed line).\label{fig:overlap_conf_am_BH14int}}
\end{figure*}

While formally distinct, these two approaches yield essentially the
same results quite irrespective of whether the latus rectum and
eccentricity are held fixed in (Q)BL or in QI coordinates. Because of
this, hereafter we will discuss only the results obtained when fixing
$p_{_{\rm (Q)BL}}$ and $e_{_{\rm (Q)BL}}$.

Figure~\ref{fig:overlap_conf_am_BH3}, in particular, shows the overlap
between waveforms produced in spacetime $A$ by {\sl external} orbits
and waveforms produced in a Kerr spacetime with mass $M_{_{\rm
    Kerr}}+\delta M=M_{_{\rm
    tot}}+\delta M$ and spin $J_{_{\rm Kerr}}+\delta J=J_{_{\rm tot}}+\delta J$ by orbits with
the same $p_{_{\rm (Q)BL}}$ and $e_{_{\rm (Q)BL}}$ and the same orbital frequencies.  As in
Fig.~\ref{fig:overlap_conf_pe_BH3}, the different lines mark the
margins of the relevant regions of the $(p_{_{\rm QBL}},\,e_{_{\rm QBL}})$ plane, with the blue
dashed line representing the outer ``edge of the torus'', the red
solid line representing the innermost stable bound orbits for a Kerr
spacetime with mass $M_{_{\rm Kerr}}=M_{_{\rm tot}}$ and spin $J_{_{\rm Kerr}}=J_{_{\rm tot}}$ and the
black dot-dashed line limiting the regions of the $(p_{_{\rm QBL}},\,e_{_{\rm QBL}})$ plane
where bound stable orbits have been studied.  Note the very close
match between the two waveforms, with an overlap ${\cal O}> 0.95$ in
essentially all of the relevant regions of the $(p_{_{\rm QBL}},\,e_{_{\rm QBL}})$ plane.
This is a clear indication that a confusion problem is present for
LISA measurements over a timescale below or comparable to the dephasing time.

Figures~\ref{fig:overlap_conf_am_BH14ext} and
\ref{fig:overlap_conf_am_BH14int} provide complementary information
for spacetime $B$, with the first one referring to {\sl external}
orbits and the second one to {\sl internal} ones (the meaning of the lines appearing in this figures is the same as in 
figures~\ref{fig:overlap_conf_pe_BH14ext} and
\ref{fig:overlap_conf_pe_BH14int}). In both cases it is
apparent that the overlap is always very large. The only exceptions are the 
internal orbits with the largest allowed eccentricities, for which the overlap decreases slightly,
and a very small set of orbits very close to the torus, for which Eqs.~\eqref{eq:same_per_3}--\eqref{eq:same_per_4}
  have no solutions (these orbits correspond to the blank regions inside 
the black dot-dashed line in figures~\ref{fig:overlap_conf_am_BH14ext} and
\ref{fig:overlap_conf_am_BH14int}). 

In summary, the overlap computed in the two spacetimes by varying the
mass and spin of the black hole reveals that a LISA observation
carried out over a timescale below or comparable to the dephasing time would not
allow an observer to distinguish between a Kerr and a non-pure Kerr spacetime,
even in the case in which the orbital parameters of the small body,
such as the the latus rectum and the eccentricity, were known through
astronomical observations.

\begin{figure*}[t]
\includegraphics[width=8cm, angle=-90]{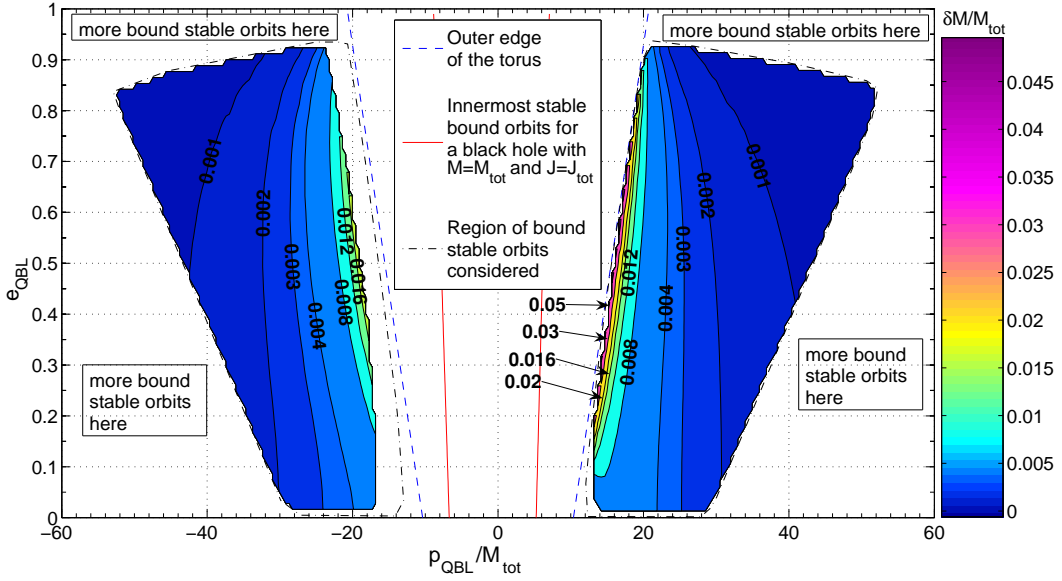}
\caption{Relative mass correction $\delta M/M_{_{\rm Kerr}}=\delta M/M_{_{\rm tot}}$ in the
  regions of the $(p_{_{\rm QBL}},\,e_{_{\rm QBL}})$ plane where the overlap plotted in
  Fig.~\ref{fig:overlap_conf_am_BH14ext} is above 0.95. Note that far
  from the system $\delta M/M_{_{\rm tot}}$ approaches zero, as
  one would expect.\label{fig:confusion_m_BH14ext}}
\vskip 0.25cm
\end{figure*}
\begin{figure*}
\vskip 0.25cm
\includegraphics[width=8.2cm, angle=-90]{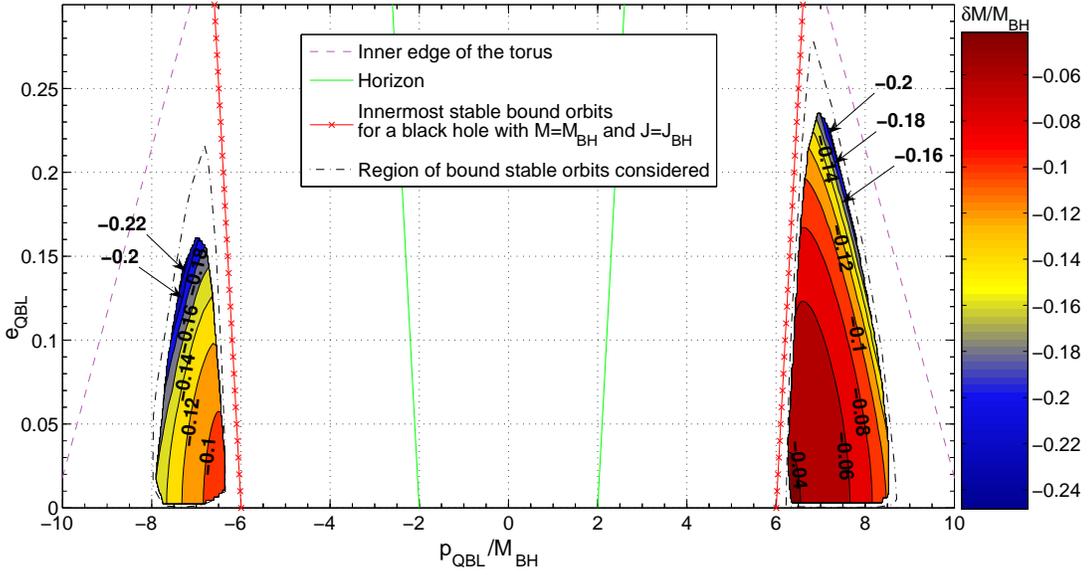}
\caption{The same as in Fig.~\ref{fig:confusion_m_BH14ext} but for
  {\sl internal orbits}. In this case the deviations are computed as
  $\delta M/M_{_{\rm Kerr}}=\delta M/M_{_{\rm BH}}$ 
  in the regions of the $(p_{_{\rm QBL}},\,e_{_{\rm QBL}})$ plane where the overlap plotted in
  Fig.~\ref{fig:overlap_conf_am_BH14int} is above 0.95.
\label{fig:confusion_m_BH14int}}
\end{figure*}

A simple explanation of why the overlap is always so large when
calculated by varying the mass and spin of the Kerr black hole is
already illustrated in Fig.~\ref{fig:tone1}. This shows that the
waveform obtained in this way captures not only the proper orbital
frequencies, but also the overall ``form'' of the signal, which is
most sensitive to the values of the latus rectum and of the
eccentricity of the orbit (cf. the solid black line and the brown
circles in Fig.~\ref{fig:tone1}).

The difficulty of distinguishing a Kerr
spacetime from a non-pure Kerr one can also be expressed in terms of
the mass $M_{_{\rm Kerr}} + \delta M$ and spin $J_{_{\rm Kerr}} +
  \delta J$ that would be measured by an observer analyzing a gravitational wave from a black hole-torus system 
  with pure Kerr templates.
  The corrections $\delta M$ and $\delta J$ are those appearing in
  Eqs.~\eqref{eq:same_per_3}--\eqref{eq:same_per_4} and have been
  computed to determine the overlaps presented in this section. If
  they are small and slowly varying, it is hard to imagine a way 
  in which the non-pure Kerr spacetime could be
  distinguished from a pure Kerr one, even with the help of additional
  astronomical observations. Conversely, if these corrections are large or rapidly varying it is possible
  that additional astronomical information on the system or an analysis of snapshots of the waveform taken at different times
  could be used to determine that the source
  is not an isolated Kerr black hole and therefore lessen the confusion problem we find in our analysis.

A synthesis of these corrections for the determination of the mass of
the black hole in the case of spacetime $B$ is presented in
Fig.~\ref{fig:confusion_m_BH14ext} and
Fig.~\ref{fig:confusion_m_BH14int}, with the first one showing the
relative error $\delta M/M_{_{\rm Kerr}}=\delta M/M_{_{\rm tot}}$ in the regions of the
$(p_{_{\rm QBL}},\,e_{_{\rm QBL}})$ plane where the overlap plotted for {\sl external} orbits is
above 0.95, and the second one showing the corresponding quantity ($\delta M/M_{_{\rm Kerr}}=\delta M/M_{_{\rm BH}}$) for
{\sl internal} orbits.

Clearly, the corrections are very
small and slowly varying in almost all of the relevant space of
parameters for {\sl external} orbits, meaning that an observer could not detect the presence of
the torus using only these orbits. On the other hand, an observer could measure rather accurately
the total mass of the system. Note in particular that the correction 
$\delta M/M_{_{\rm tot}}$ goes to zero far from the system, as one would expect.

This situation is only slightly different for
{\sl internal} orbits, for which the correction increases to some percent:
using internal orbits an observer could measure quite accurately the mass of the central black hole. Note therefore that a combination
of observations of internal orbits (giving an estimate for $M_{_{\rm BH}}$) and external orbits (giving an estimate for $M_{_{\rm tot}}$) could hint at the presence of a torus around the central black hole.

Similar behaviour has also been found for spacetime $A$.
Because no internal bound stable orbits
are present, an observer could not measure the individual masses of the black hole and the torus, whereas
he could measure accurately the total mass of the system. 
In fact, the corrections are always very small with $\vert\delta
M/M_{_{\rm Kerr}}\vert=\vert\delta
M/M_{_{\rm tot}} \vert\lesssim 0.02$; again, the  correction 
$\delta M/M_{_{\rm tot}}$ goes to zero far from the system, as one would expect.
Note that due to the absence of internal orbits in this spacetime and to the smallness and slow variations of $\delta
M/M_{_{\rm tot}}$ it extremely difficult to distinguish spacetime $A$ from a pure Kerr spacetime.

\begin{figure*}
\includegraphics[width=8.2cm, angle=-90]{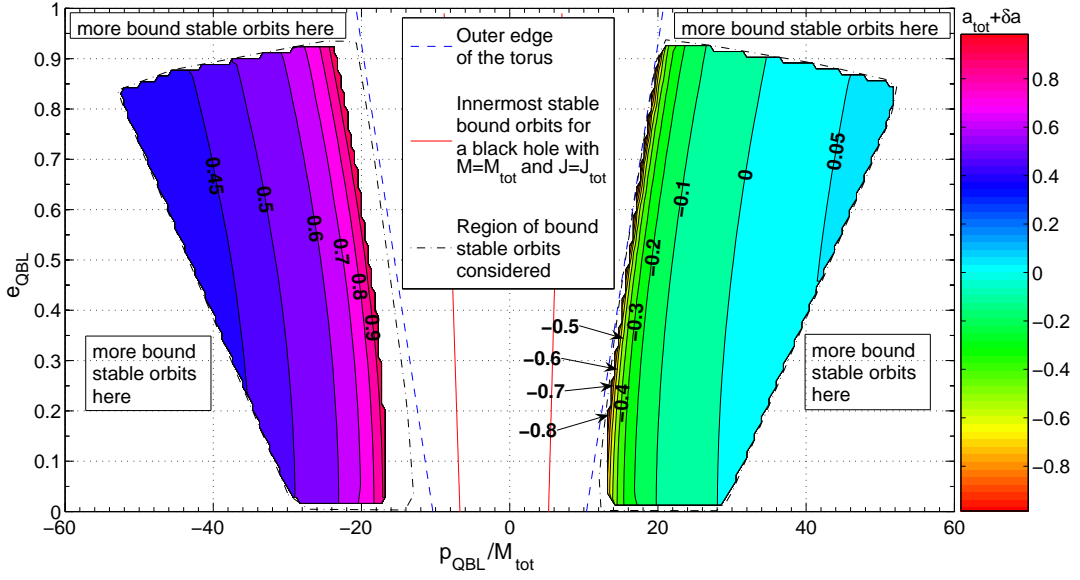}
\caption{Variations of the spin $a_{_{\rm Kerr}}+\delta a\equiv a_{_{\rm tot}}+\delta a=(J_{_{\rm
      tot}}+\delta J)/(M_{_{\rm tot}}+\delta M)^2$ in the regions of the $(p_{_{\rm QBL}},\,e_{_{\rm QBL}})$
  plane where the overlap plotted in
  Fig.~\ref{fig:overlap_conf_am_BH14ext} is above 0.95. We recall that
  for {\sl external} orbits in spacetime $B$, we have 
  $a_{_{\rm Kerr}}=a_{_{\rm tot}}=0.224$
  (cf. Table~\ref{table:spacetimes_data}).\label{fig:confusion_a_BH14ext}}
\vskip 0.25cm
\end{figure*}
\begin{figure*}
\vskip 0.25cm
\includegraphics[width=8.2cm, angle=-90]{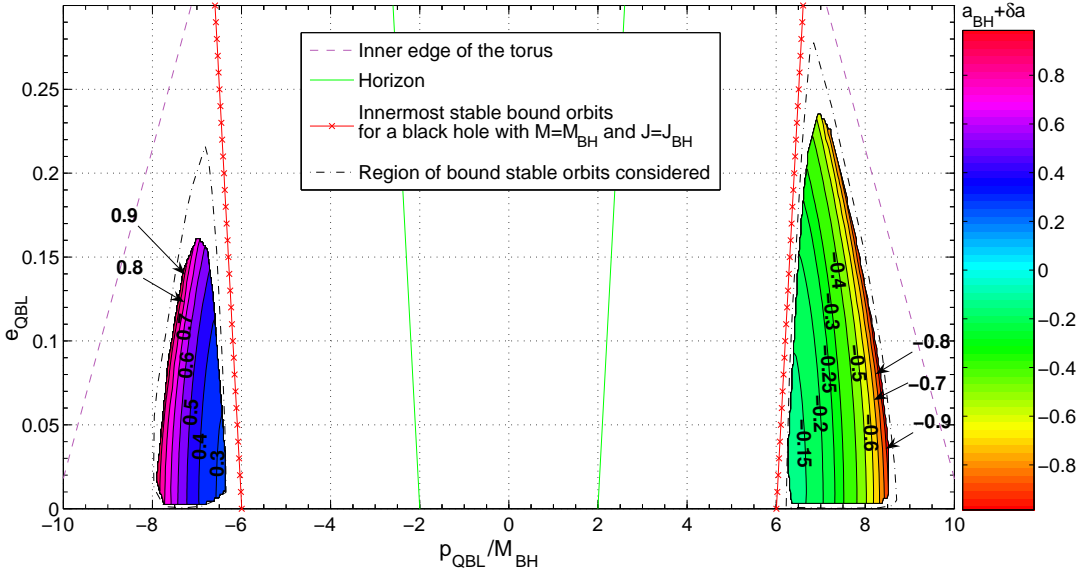}
\caption{The same as in Fig.~\ref{fig:confusion_a_BH14ext} but for
  {\sl internal} orbits. In this case the corrections are computed as
  $a_{_{\rm Kerr}}+\delta a=a_{_{\rm BH}}+\delta a\equiv(J_{_{\rm BH}}+\delta J)/(M_{_{\rm
      BH}}+\delta M)^2$  in the regions of the $(p_{_{\rm QBL}},\,e_{_{\rm QBL}})$
  plane where the overlap plotted in
  Fig.~\ref{fig:overlap_conf_am_BH14int} is above 0.95.  We recall that for {\sl internal} orbits
  in spacetime $B$, we have $a_{_{\rm Kerr}}=a_{_{\rm BH}}=-1.74\times10^{-3}$
  (cf. Table~\ref{table:spacetimes_data}).
\label{fig:confusion_a_BH14int}}
\end{figure*}

Information complementary to the one given by
the mass correction $\delta M$ is offered 
by the spin correction $\delta J$. In particular, for spacetime $A$ the correction
$\delta a$ defined by
$\delta a\equiv(J_{_{\rm Kerr}}+\delta J)/(M_{_{\rm Kerr}}+\delta M)^2-a_{_{\rm Kerr}}$
(with $a_{_{\rm Kerr}}=J_{_{\rm Kerr}}/M^2_{_{\rm Kerr}}=0.728$) can be readily calculated to be 
$|\delta a/a_{_{\rm Kerr}}|\lesssim 0.065$, going to zero, as one would expect, far from the system.
This means that an observer could accurately measure the total spin of the black hole-torus system although,
due to the absence of internal orbits in this system and to the slow variations of $\delta a$, a measurement of the individual spins of the torus and the black hole or even a simple detection of the torus seems unfeasible. 

Spacetime $B$ is considered in figures~\ref{fig:confusion_a_BH14ext}-\ref{fig:confusion_a_BH14int}, in which
we report the quantity $a_{_{\rm Kerr}}+\delta a
\equiv(J_{_{\rm Kerr}}+\delta J)/(M_{_{\rm Kerr}}+\delta M)^2$ for external (with $a_{_{\rm Kerr}}=J_{_{\rm Kerr}}/M^2_{_{\rm Kerr}}=0.224$) 
and internal  (with $a_{_{\rm Kerr}}=J_{_{\rm Kerr}}/M^2_{_{\rm Kerr}}=-1.74\times10^{-3}$) orbits,
respectively. As can be seen, the corrections $\delta a$ are, in both cases, rather large and rapidly varying: an observer could probably
distinguish this spacetime from a pure Kerr one using estimates of the spin obtained by analyzing the waveform at different times, but
would have little chance to measure the spin of the central black hole correctly and should consider orbits very far from the system
in order to achieve accurate measurements of the total spin. This was to be expected, since spacetime $B$
has a large ratio $J_{_{\rm Torus}}/J_{_{BH}}$, which causes the quadrupole parameter $\epsilon$ to be large (cf. Table~\ref{table:spacetimes_data}).

Before concluding this section, it is worth commenting
on how robust and generic these results are. While we believe they
represent the first attempt to model consistently the
gravitational-wave emission from spacetimes that deviate considerably
for Kerr due to the presence of matter, the approach followed here has the obvious limitation of
neglecting radiation-reaction effects and thus of considering
waveforms only over a dephasing time which is typically of days or
weeks. It is therefore possible, if not likely, that considering
waveforms over a timescale comparable with LISA's planned lifetime
(i.e. 3--5 years) would lower the overlaps computed here and thus
reduce the impact of a confusion problem.

As already mentioned, a simple way to
include radiation reaction would consist of using the
adiabatic approximation and thus considering motion along a geodesic with
slowly changing parameters. In particular, approximate (``kludge'') expressions
for the fluxes $\dot E$, $\dot{L}_z$ and $\dot{Q}$ in Kerr have been derived using post-Newtonian
expansions~\cite{ryan_fluxes,kludge_first_paper}, recently corrected using fits to the fluxes computed rigorously with the Teukolsky
formalism~\cite{GG_kludge_fluxes}. Likewise, it may be possible to adopt similar strategies
in non-Kerr spacetimes. For instance, Cutler \& Barack~\cite{quasi_kerr_cutler} recently proposed including
radiation reaction in quasi-Kerr spacetimes by using post-Newtonian
fluxes in which the leading-order effect of the quadrupole of the
spacetime is taken into account, potentially eliminating the confusion problem. 
Nevertheless, it is still unclear at this stage whether post-Newtonian fluxes 
will be a good approximation for our spacetimes, where the parameters $\epsilon$ and $a$ can be
$O(1)$. We recall, indeed, that using post-Newtonian fluxes is not
always a good approximation even in Kerr spacetimes and that the most accurate ``kludge'' fluxes for Kerr~\cite{GG_kludge_fluxes}
are certainly based on post-Newtonian expansions, but are also corrected using fits to
rigorous Teukolsky-based fluxes.

We also note than even with the radiation-reaction included, a
``confusion'' problem might in principle still be present, at least for equatorial
orbits. In fact, requiring the equality
of the $r$- and $\phi$-frequencies fixes only two of the four free
parameters characterizing the geodesic, $p$, $e$, $M_{_{\rm Kerr}}$
and $J_{_{\rm Kerr}}$, while the remaining two could be used to obtain the equality
of the time derivatives of the $r$- and $\phi$-frequencies
at the initial time
[$\dot{\omega}^{^{\rm BH+Torus}}_r(t_0)=\dot{\omega}^{^{\rm Kerr}}_r(t_0)$,
$\dot{\omega}^{^{\rm BH+Torus}}_\phi(t_0)=\dot{\omega}^{^{\rm Kerr}}_\phi(t_0)$], which could ensure, at least initially,
a similar evolution under radiation reaction for the two
waveforms.

Besides inclusion of radiation reaction,
three other approaches to improve the estimates computed in this
paper are also worth considering. The first and most obvious one
consists of replacing the ``kludge'' waveforms with more rigorous
waveforms, solutions of Eqs.~\eqref{eq:box_h} and~\eqref{eq:euler},
possibly neglecting the fluid perturbations appearing on the
right-hand-side of Eq.~\eqref{eq:box_h} (the latter could be a rather
good approximation for orbits far enough from the torus.). Doing this
in practice is certainly not trivial since Eqs.~\eqref{eq:box_h} have
been solved only for a Schwarzschild
spacetime~\cite{barack} so far. The second improvement is simpler and
involves considering tori which are not as compact and close to the
black hole as the ones studied here, but are instead a better
approximation of those observed around SMBHs in
AGNs. Finally, the third possible improvement involves the extension
of the present analysis to non-equatorial orbits. While this is
more complicated as one cannot require the strict
equality of the orbital frequencies [in contrast to Kerr,
  Eqs.~\eqref{eq:geo_t}--\eqref{eq:geo_theta} indicate that in general
  the $r$-, $\theta$- and $\phi$-motions are not periodic in the time
  coordinate $t$], the motions in the $r$-, $\theta$- and
$\phi$-directions are almost periodic if the torus is not too massive
and hence the present analysis can be extended straightforwardly in
terms of these almost-periodic motions.

%-----------------------------------------------
\section{Conclusions}\label{sec:conclusion}
%-----------------------------------------------

EMRIs are expected to be among the most important sources for LISA and,
besides mapping accurately the spacetime around SMBHs, they might also shed light on the distribution of matter around
them. We have here studied EMRIs and the corresponding
gravitational-wave emission in spacetimes that are highly-accurate
numerical solutions of the Einstein equations and consist
of an SMBH and a compact torus with comparable mass and spin. We
underline that the tori considered here \textit{do not} represent a
model for the accretion disks in AGNs but, rather, are a
phenomenological model for a compact source of matter close to the
SMBH. Our goal in this paper has therefore been that of maximizing the impact of this matter on the waveforms,
investigating whether gravitational-wave observations will be able to reveal its
presence. This hypothetical matter source, even if it exists, 
may not be detectable otherwise, being too close to the central SMBH and possibly ``dark''.

Using the semi-relativistic approach proposed in
Ref.~\cite{quasi_kerr}, we have compared kludge waveforms produced by
equatorial orbits in non-pure Kerr spacetimes with waveforms produced
by equatorial orbits in Kerr spacetimes having the same mass and
spin as the non-pure Kerr spacetimes. Because they are produced by purely
geodetic motion, these waveforms are valid only over a rather short
``dephasing'' timescale. Overall, we find that waveforms produced by
orbits having the same latus rectum and eccentricity $p$ and $e$ are
considerably different throughout the whole space of parameters
$(p,\,e)$. On the other hand, comparisons of waveforms produced by
(equatorial) orbits having the same $r$- and $\phi$-frequencies, with
this condition being achieved by changing the latus rectum and
eccentricity of the orbits in the Kerr spacetime, produce overlaps
${\cal O}>0.95$ for orbits far from the black hole-torus system, hence
pointing out a confusion problem. This overlap decreases rapidly as
one considers orbits which are close to the torus, indicating that in the strong-field region no confusion problem
is present. Finally, if the equality of the $r$- and
$\phi$-frequencies is obtained by changing the mass and spin of the Kerr
spacetime while maintaining fixed the latus rectum and the
eccentricity of the orbit, the resulting overlaps are very high, with ${\cal
  O}>0.99$ for essentially all of the orbital parameters $p$ and $e$,
indicating a confusion problem that is less severe only 
for a few orbits very close to the torus.

This confusion problem in the mass and the spin 
might therefore be more serious than the one
involving latus rectum and eccentricity. Stated differently, an
observer analyzing below the dephasing timescale a gravitational
waveform produced by an EMRI in a black hole-torus system would not
be able to distinguish it from one produced in a pure Kerr
spacetime. This observer would therefore associate the EMRI to a Kerr
SMBH whose mass and spin would however be estimated incorrectly.

While these results represent the first attempt to model consistently
the gravitational-wave emission from spacetimes that deviate
considerably for Kerr, the approach followed here is based on four
approximations, namely: \textit{i)} the use of kludge waveforms in
place of ones that are consistent solutions of the Einstein equations;
\textit{ii)} the use of a cut-off at the dephasing time beyond which
radiation-reaction effects can no longer be ignored; \textit{iii)} the
restriction to purely equatorial orbits; \textit{iv)} the use of tori
that are very compact and close to the black hole. Work is now in
progress to relax one or more of these approximations, with the
expectation that this will lead to a less serious confusion problem.
\begin{acknowledgments}
It is a pleasure to thank S. Babak, M. Colpi, K. Glampedakis,
J. Miller, E. Poisson, T. Sotiriou and P. Ullio for fruitful and enlightening
discussions. EB would like to thank T. Bogdanovic for useful comments during the 6th LISA Symposium.
EB acknowledges support also from the European Network
of Theoretical Astroparticle Physics ILIAS/N6 (contract number
RII3-CT-2004-506222). DP was supported by the DFG grant SFB/Transregio 7, ``Gravitational Wave Astronomy''.
\end{acknowledgments}

%-----------------------------------------------
\appendix*
\section{From the Einstein equations to the semirelativistic approach}
\label{appendix_A}
%-----------------------------------------------

Although the main motivation for the semirelativistic approach we use in this paper is the the surprising agreement that ``kludge'' waveforms show in Kerr with the rigorous waveforms computed using the Teukolsky formalism~\cite{babak_kludge}, one can also try to make sense of it using the Einstein equations.

We start by
rewriting the Einstein equations in a more convenient form in which we
isolate the perturbation as ~\cite{MTW,Will}
\eq 
\bar{H}^{\mu \nu} \equiv \eta^{\mu\nu} -
(-{\widetilde g})^{1/2} {\widetilde g}^{\mu \nu}\;,
\label{hdefinition}
\eeq 
where $\eta^{\mu\nu}$ is the Minkowski metric. 
Since far from the
source the spacetime reduces to Minkowski plus a small perturbation,
i.e. ${\boldsymbol g} = {\boldsymbol \eta}$, the first-order
perturbations there coincide with ${\bar{\boldsymbol H}}$, i.e.
$\bar{H}^{\mu \nu}=\bar{h}^{\mu \nu}+{\cal O}(m/M)^2$, with
$\bar{h}^{\mu \nu}$ being the trace-reversed potentials defined in
Eq.~\eqref{rev_trace}.

If we now restrict our attention to a region of the spacetime where it is possible to choose the harmonic gauge
\eq
\partial_\beta\bar{H}^{\alpha \beta} = 0\;
\label{harmonic}
\eeq
(this is always possible far enough from the source),
the \textit{full} Einstein equations give~\cite{Will}
\eq
\Box_{\rm flat} \bar{H}^{ \alpha \beta } = -16 \pi {\tau}^{ \alpha \beta } \; ,
\label{relaxed}
\eeq
where $\Box_{\rm flat} \equiv \eta^{\mu\nu}\partial_\mu\partial_\nu $
is the flat-spacetime wave operator.  The right-hand
side is given by the effective stress-energy pseudotensor 
\eq
\tau^{\alpha\beta} = (-{\widetilde g}){\widetilde T}^{\alpha\beta} +
	(16\pi)^{-1} \Lambda^{\alpha\beta} \;,
\label{effective}
\eeq
where $\Lambda^{\alpha\beta}$ is given by
\begin{equation}
\Lambda^{\alpha \beta}
   = 16\pi (-{\widetilde g}) t_{_{\rm LL}}^{\alpha \beta }
   +  ( \bar{H}^{\alpha \mu},_{\nu} \bar{H}^{\beta \nu},_{\mu}
                  - \bar{H}^{\alpha \beta},_{\mu \nu} \bar{H}^{\mu \nu} ) \; ,
\label{nonlinear}
\end{equation}
and $ t_{_{\rm LL}}^{\alpha \beta }$ is the Landau-Lifshitz pseudotensor
\begin{eqnarray}
&& 16 \pi (-{\widetilde g})t_{_{\rm LL}}^{\alpha \beta } \equiv
 {\widetilde g}_{\lambda\mu}{\widetilde g}^{\nu\rho}{ \bar{H}^{\alpha\lambda}}_{,\nu}{ \bar{H}^{\beta\mu}}_{,\rho}\\
&&+\frac{1}{2}
 {\widetilde g}_{\lambda\mu}{\widetilde g}^{\alpha\beta}{ \bar{H}^{\lambda\nu}}_{,\rho}{ \bar{H}^{\rho\mu}}_{,\nu}
- 2{\widetilde g}_{\mu\nu}{\widetilde g}^{\lambda (\alpha}{ \bar{H}^{\beta )\nu}}_{,\rho}{ \bar{H}^{\rho\mu}}
_{,\lambda}\nonumber
\\
&&+ \frac{1}{8}
(2{\widetilde g}^{\alpha\lambda}{\widetilde g}^{\beta\mu}-{\widetilde g}^{\alpha\beta}{\widetilde g}^{\lambda\mu})
(2{\widetilde g}_{\nu\rho}{\widetilde g}_{\sigma\tau}-{\widetilde g}_{\rho\sigma}{\widetilde g}_{\nu\tau})
{ \bar{H}^{\nu\tau}}_{,\lambda}{ \bar{H}^{\rho\sigma}}_{,\mu}  \;.\nonumber
\label{landau}
\end{eqnarray}
Because of the gauge condition~\eqref{harmonic}, the source term of Eq.~\eqref{relaxed}
satisfies the conservation law
\begin{equation}
{{\tau}^{\alpha \beta}}_{, \beta} = 0 \; ,
\label{conservation}
\end{equation}
which is equivalent to the equations of motion of the matter

\eq\label{conservation2} \widetilde{\nabla}_\beta{\widetilde T}^{\alpha\beta}=0\;. \eeq

\noindent Combining then Eqs.~\eqref{relaxed} and~\eqref{conservation}, in the slow
motion approximation one easily gets the usual quadrupole formula
(see Ref.~\cite{MTW} for details): 
\begin{align}&\bar{H}^{ij}(\vec{x},t)=\frac{2}{r}\left[\frac{d^2 I^{ij}}{dt'^2}\right]_{t'=t-r}\;,\label{quadrupole_formula}\\
&I^{ij}(t')=\int\tau^{00}(\vec{x}',t')x'^i x'^j{\rm d}^3x'\;,\label{quadrupole_def} \end{align}
where $r^2\equiv\vec{x}\cdot\vec{x}$.
Note that one can easily relax the slow motion assumption by
including the octupole terms~\cite{Bekenstein} or even all the higher
order multipoles (the formula is due to Press~\cite{press}).

Eq.~\eqref{quadrupole_formula} clearly does not allow one to compute  $\bar{H}^{ij}$ directly,
because its right hand side depends on $ \bar{H}^{\alpha\beta}$ [cf. Eq.~\eqref{effective}].
The semirelativistic approximation
consists indeed in \textit{pretending} that $\bar{\boldsymbol H}$ is ``small'':
making this assumption, one can neglect, in the expression~\eqref{effective} for the effective stress-energy tensor
$\tau^{\alpha\beta}$, the terms quadratic in
$\bar{H}^{\alpha\beta}$ and the terms in which $\bar{H}^{\alpha\beta}$ is multiplied by the mass
$m$ of the small body. In addition, the semirelativistic approximation also neglects all the terms involving the
stress-energy tensor of the fluid: with these assumptions, $\tau^{\alpha\beta}$ 
can be written as
\begin{align}
&\tau^{00}(\vec{x},t)=m\,\gamma(t)\,\delta^{(3)}(\vec{x}-\vec{z}(t))\;,\label{kl1}\\
&\tau^{0i}(\vec{x},t)=m\,\gamma(t)\,\dot{z}^i(t)\,\delta^{(3)}(\vec{x}-\vec{z}(t))\;,\label{kl2}\\
&\tau^{ij}(\vec{x},t)=m\,\gamma(t)\dot{z}^i(t)\,\dot{z}^j(t)\,\delta^{(3)}(\vec{x}-\vec{z}(t))\;,\label{kl3}\\
&\gamma\equiv(1-\delta_{ij}\dot{z}^i\dot{z}^j)^{-1/2}\;,\nonumber
\end{align}
where the dot indicates the derivative with respect to the coordinate time $t$ and the trajectory $z^i(t)$ of the small body is
obtained by solving the geodesic equations, which
are indeed contained in Eq.~\eqref{conservation2}. Note that Eqs.~\eqref{kl1}-\eqref{kl3} represent the stress-energy tensor of a small body moving along the trajectory $z^i(t)$
in a Minkowski spacetime, which constitutes exactly the assumption on which kludge waveforms are based.
In particular, the quadrupole moment~\eqref{quadrupole_def} reduces, in the slow motion approximation, to its 
textbook version $I^{ij}(t)=m z^i(t) z^j(t)$, while analogous simplifications
happen for the octupole and Press formulas (see Ref.~\cite{babak_kludge} for details). 

Having calculated $\bar{H}^{ij}\approx \bar{h}^{ij}$, it is then a trivial
task to project out the gauge invariant transverse traceless
perturbations $h_+$ and $h_\times$ at infinity (see for instance
Refs.~\cite{MTW,babak_kludge} for details).

\end{document}